\documentclass[journal,twoside,web]{ieeecolor}

\usepackage{generic}
\usepackage{cite}
\usepackage{amsmath,amssymb,amsfonts}
\usepackage{algorithmic}
\usepackage{graphicx}
\usepackage{xcolor}
\usepackage{algorithm,algorithmic}
\usepackage{booktabs}
\usepackage{comment}
\usepackage{colortbl}
\usepackage{bbding}
\usepackage{gensymb}
\usepackage{soul}
\usepackage{orcidlink}

\usepackage{graphicx}
\usepackage{float}
\usepackage{dblfloatfix}
\usepackage{multirow}

\usepackage{textcomp}
\def\BibTeX{{\rm B\kern-.05em{\sc i\kern-.025em b}\kern-.08em
    T\kern-.1667em\lower.7ex\hbox{E}\kern-.125emX}}
\begin{document}
\title{MedMimic: A Physician-Inspired Multimodal Fusion Framework for Early Diagnosing Fever of Unknown Origin}
\author{Minrui Chen, \IEEEmembership{Student Member, IEEE} \textsuperscript{\orcidlink{0000-0002-5845-3200}}, Yi Zhou\textsuperscript{\orcidlink{0009-0005-0826-4224}}, Huidong Jiang\textsuperscript{\orcidlink{0009-0003-3859-0536}}, Yuhan Zhu\textsuperscript{\orcidlink{0009-0001-8614-0535}}, Guanjie Zou\textsuperscript{\orcidlink{0009-0005-1687-5316}}, Minqi Chen\textsuperscript{\orcidlink{0009-0009-6960-8667}}, Rong Tian\textsuperscript{\orcidlink{0000-0002-5191-9004}}, and Hiroto Saigo\textsuperscript{\orcidlink{0000-0001-5314-5367}}
\thanks{This work was supported in part by the National Natural Science Foundation of China under Grant Number 81971653 (Corresponding authors: Rong Tian, Hiroto Saigo).}
\thanks{This work involved human subjects and/or animals. All ethical and experimental procedures were approved by the Medical Ethics Committee of West China Hospital (Application No. 2023-954).}
\thanks{Minrui Chen and Yi Zhou contributed equally to this work.}
\thanks{Minrui Chen is with the Department of Information Science and Technology, Kyushu University, Fukuoka, Japan, e-mail: chen.minrui.978@s.kyushu-u.ac.jp.}
\thanks{Yi Zhou is with the Department of Nuclear Medicine, West China Hospital, Sichuan University, Chengdu, China, e-mail: 1614693769@wchscu.cn.}
\thanks{Huidong Jiang is with the Department of Computer Science, Institute of Science Tokyo, Yokohama, Japan, and the Center for Advanced Intelligence Project, RIKEN, Tokyo, Japan, e-mail: jiang.h.af@m.titech.ac.jp.}
\thanks{Yuhan Zhu is with the Department of Computational Biology and Medical Sciences, University of Tokyo, Tokyo, Japan, e-mail: yuhanzhu@g.ecc.u-tokyo.ac.jp.}
\thanks{Guanjie Zou is with the Department of Computational Biology and Medical Sciences, University of Tokyo, Tokyo, Japan, e-mail: 8923236439@edu.k.u-tokyo.ac.jp.}
\thanks{Minqi Chen is with the Department of Thoracic Surgery, West China Hospital, Sichuan University, Chengdu, China, and Institute of Thoracic Oncology, West China Hospital, Sichuan University, Chengdu, China, e-mail: 2019141230152@stu.scu.edu.cn.}
\thanks{Rong Tian is with the Department of Nuclear Medicine, West China Hospital, Sichuan University, Chengdu, China, e-mail: tianrong@wchscu.cn.}
\thanks{Hiroto Saigo is with the Department of Electrical Engineering and Computer Science, Kyushu University, Fukuoka, Japan, e-mail: saigo@inf.kyushu-u.ac.jp.}
}

\maketitle

\begin{abstract}
Fever of unknown origin (FUO) presents a major diagnostic challenge, often necessitating extensive evaluations. Although integrating clinical features with \(^{18}\)F-FDG PET/CT imaging has enhanced diagnostic accuracy, conventional feature extraction can lack generalizability, and limited annotated datasets restrict deep learning. To address these issues, we propose Medical Mimicry (MedMimic), a framework that leverages pre-trained models—including DINOv2, Vision Transformer, and ResNet-18—to transform high-dimensional PET/CT data into semantically meaningful feature tensors. A learnable self-attention-based fusion network then integrates imaging features with clinical data, producing compact yet discriminative representations for downstream classification. Our dataset from Sichuan University West China Hospital included 607 consecutive FUO patients (January 2017 to December 2023), with 416 admissions remaining after exclusions. Specifically, we decomposed the diagnostic challenge into seven classification tasks reflecting common clinical etiologies, enabling more targeted model evaluation. Our multimodal fusion classification network (MFCN) achieved macro-averaged area under the receiver operating characteristic curve (macro-AUROC) values of 0.8654 to 0.9291 across these tasks, outperforming competing machine learning (ML) and single-modality deep learning (DL) approaches. Ablation studies confirmed the efficacy of each MFCN component, while five-fold cross-validation demonstrated consistent performance. By harnessing the strengths of pre-trained large models and deep learning, MedMimic offers a novel perspective and practical solution for disease classification. These findings underscore the potential of leveraging large-scale, pre-trained architectures in complex clinical scenarios, bridging the gap between imaging data and clinical decision-making. Future work will explore external validations in multi-center settings and the extension of MedMimic to other diagnostic domains, emphasizing its versatility and clinical value.
\end{abstract}


\section{Introduction}
\label{I}

\begin{figure}[htbp]
  \centering
  \includegraphics[width=0.5\textwidth]{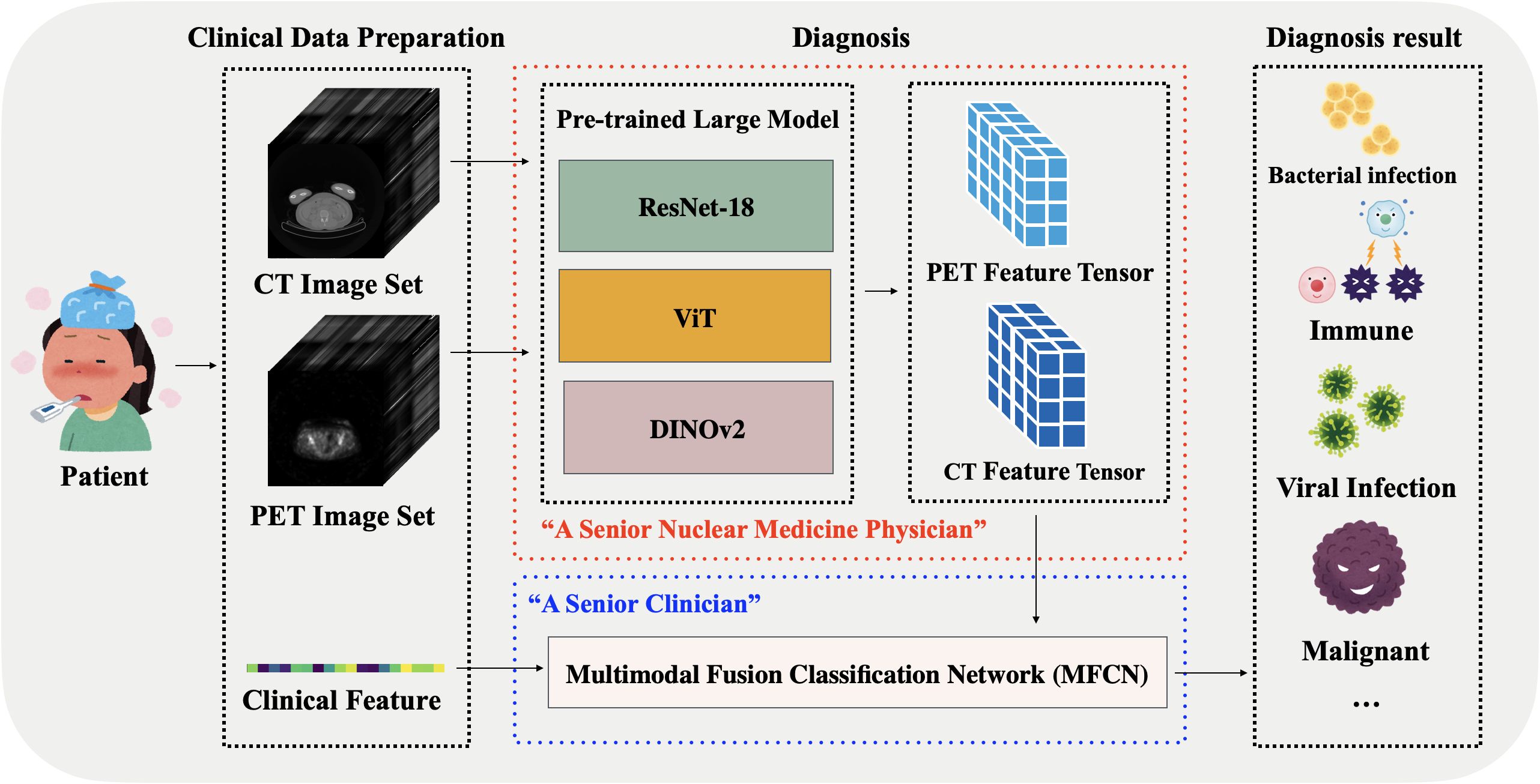}
  \caption{A diagnostic process of the proposed MedMimic framework.}
  \label{MedMimic: Multimodal Feature Fusion for FUO Diagnosis}
\end{figure}

\IEEEPARstart{F}{ever} of unknown origin (FUO) has posed a clinical challenge for over a century, defined as a prolonged fever exceeding 38.3\,\degree\text{C} lasting more than three weeks without an identified cause after extensive diagnostic evaluations \cite{cabot1907three}\cite{wright2024fever}. FUO encompasses a wide range of etiologies, including infections, autoimmune disorders, and malignancies, with many cases remaining undiagnosed even after thorough assessments \cite{durack1991fever}. Empirical treatments initiated without a definitive diagnosis frequently result in unnecessary side effects or obscure the underlying cause, complicating both diagnosis and management \cite{bryan2007fever}\cite{cunha2015fever}.

The advent of machine learning (ML) in medical research has introduced new possibilities for supporting FUO diagnosis. For example, Xu et al. \cite{xu2022fever} developed a logistic regression model based on five clinical and inflammatory markers to predict bloodstream infections in FUO patients. Similarly, Yan et al. \cite{yan2021fever} approached FUO diagnosis as a multi-classification task, evaluating five ML models, including LightGBM and random forests, across 18 clinical indicators, with LightGBM achieving the best performance. Wang et al. \cite{wang2023fever} proposed an interpretable hierarchical multimodal neural network framework, integrating medical knowledge with multimodal data to enhance diagnostic precision.

Among advanced diagnostic tools, \textsuperscript{18}F-FDG PET/CT has shown considerable promise. Minamimoto \cite{minamimoto2022fever} demonstrated its effectiveness in improving diagnostic accuracy and patient management when combined with comprehensive clinical evaluations. Chen et al. \cite{chen2023fever} further enhanced its diagnostic utility by introducing a scoring system to distinguish lymphoma from benign causes in FUO patients with lymphadenopathy. Additional studies \cite{georea2020fever, palestro2023fever} highlight \textsuperscript{18}F-FDG PET/CT's sensitivity in identifying FUO causes, aiding diagnosis in 70\% of cases by detecting active disease or guiding biopsy site selection. As a whole-body imaging modality, PET/CT has proven to be more effective than older methods such as Gallium-67 scintigraphy in pinpointing fever causes and guiding treatment \cite{palestro2023fever}.

Despite these advances, diagnosing FUO remains challenging due to its over 200 potential underlying causes, highly diverse clinical presentations, and significant regional and temporal variations \cite{haidar2022fever}. These complexities place a considerable burden on radiologists, who often struggle to extract accurate and representative features from CT and PET scans. Furthermore, the integration of high-dimensional imaging data with low-dimensional clinical data is a significant challenge for traditional ML methods, which are not well-suited to handle such multimodal complexities.

Emerging DL methods offer a promising alternative by enabling seamless integration of medical imaging and clinical data, thereby improving diagnostic accuracy and clinical decision-making. For instance, Huang et al. \cite{huang2020CTwithDL} systematically reviewed DL approaches that combine imaging data with electronic health records (EHRs), emphasizing their potential to enhance diagnostic precision. These methods are particularly well-suited for addressing the multimodal nature of FUO diagnosis, where clinical and imaging data need to be synthesized into a cohesive diagnostic framework.

Real-world clinical data availability remains a significant barrier to advancing research in FUO. Studies \cite{kubota2021fever, kim2017fever, spernovasilis2020fever} indicate that only a small proportion of FUO patients have accessible PET/CT data, limiting the development and validation of large-scale diagnostic models. For instance, a multicenter study in Japan \cite{liu2024datasets} reported that among 128 patients who underwent PET/CT, only 92 cases resulted in successful diagnoses, highlighting the scarcity of comprehensive datasets. This shortage not only constrains the applicability of cutting-edge machine learning techniques but also underscores an urgent need for innovative strategies to bridge these data gaps and enhance the utility of advanced diagnostic tools.

Compounding these challenges is a critical shortage of nuclear medicine physicians, a pressing issue emphasized by Bluth et al. \cite{bluth2022lake}. The demand for advanced diagnostic and therapeutic imaging services is increasing, yet Scott et al. \cite{scott2024lakeofdoctor} noted the high costs and extended timelines required to train specialized professionals, creating significant barriers to workforce development. Moreover, Hricak et al. \cite{hricak2021lakeofdoctor} highlighted that the shortage of adequately trained personnel has particularly impacted resource-limited regions, where the implementation and effective utilization of nuclear medicine technologies remain suboptimal. Together, these limitations pose critical challenges to advancing the field and ensuring equitable access to life-saving diagnostic and therapeutic tools.

To address the challenges of high-dimensional and multimodal data fusion, we propose a novel diagnostic framework that leverages pre-trained models for feature extraction and integrates them with a learnable self-attention-based multimodal fusion network. Inspired by clinicians' diagnostic reasoning, the framework combines patient characteristics, laboratory results, and \(^{18}\)F-FDG PET/CT imaging data to enhance diagnostic accuracy and scalability. This approach promotes early and precise diagnosis of FUO while ensuring cost-effectiveness.

Our methodology begins with the aggregation and structuring of multimodal data into a unified pipeline. Clinical features are standardized, and imaging features are extracted using pre-trained models before being integrated into a learnable self-attention-based fusion network. This design enables diverse information sources to interact synergistically, facilitating comprehensive diagnostic reasoning.

Building upon this foundation, the proposed staged multimodal fusion framework comprises three main steps. First, clinical data preparation ensures standardized test indices for patients. Second, pre-trained models extract multi-level features from CT and PET scans, refining them into compact and discriminative representations. Finally, a learnable self-attention-based fusion network integrates clinical, CT, and PET features into a cohesive diagnostic model. This approach transforms FUO diagnosis into a multimodal classification task, significantly enhancing efficiency and interpretability.

The primary contributions of this study are as follows:
\begin{enumerate}
    \item \textbf{Novel Diagnostic Framework:} We introduce MedMimic, an innovative diagnostic framework that integrates pre-trained models for feature extraction with a learnable self-attention-based multimodal fusion network. Designed to emulate real-world clinical diagnostic reasoning, MedMimic seamlessly incorporates clinical parameters and \(\mathrm{^{18}F}\)-FDG PET/CT imaging features. This approach effectively addresses the challenges of high-dimensional data fusion, improving diagnostic accuracy while optimizing cost-effectiveness.

    \item \textbf{Learnable Self-Attention:} Our framework redefines FUO diagnosis as a multimodal classification task by bridging the gap between high-dimensional imaging features and low-dimensional clinical data. Furthermore, we design a learnable self-attention layer inspired by the clinical diagnostic process, dynamically adjusting weights to information from different modalities. This mechanism enhances the model’s ability to capture long-range dependencies and salient features.
    
    \item \textbf{Empirical Validation:} We validate our approach using a real-world dataset of 416 patients, demonstrating its capability to process high-dimensional, irregularly sampled multimodal data. Through 5-fold cross-validation, we confirm the framework’s robustness and performance, underscoring its potential to enhance FUO diagnostic accuracy and its seamless integration into clinical practice.
\end{enumerate}

The remainder of this article is organized as follows: Section~\ref{II} provides an overview of feature extraction using pre-trained models and multimodal fusion via self-attention. Section~\ref{III} introduces tensor notations and formulates the FUO diagnosis problem within a tensor-based framework. Section~\ref{IV} describes the dataset used and details the proposed methodology. Section~\ref{V} presents experimental results on FUO diagnosis, followed by an ablation study. Finally, Section~\ref{VI} concludes the paper. Moreover, the key notations used in this paper are summarized in Table~\ref{Summary of notations used in this paper.}.

\begin{table}[t]
\centering
\caption{Summary of notations used in this paper.}
\label{Summary of notations used in this paper.}
\resizebox{\columnwidth}{!}{%
\begin{tabular}{ll}
\toprule
\textbf{Notation} & \textbf{Description} \\
\midrule
$i$ & Patient index \\
$j$ & Slice index \\
$k$ & Feature extraction method index \\
$I$ & Total number of patients \\
$N$ & Number of CT slices \\
$M$ & Number of PET slices \\
$K$ & Total number of Feature extraction method \\
$n$ & Resolution of each CT slice \\
$m$ & Resolution of each PET slice \\
$a$ & Dimension of the clinical feature vector \\
$b$ & Dimension of the extracted feature vector \\
$y$ & The one-hot encoded label\\
$\mathbf{a}$ & Clinical feature vector \\
$\mathbf{f}$ & Feature vector \\
$\mathbf{A}$ & Clinical feature matrix \\
$\mathbf{X}$ & Image slice matrix\\
$\mathbf{F}$ & Feature matrix \\
$\mathcal{N}$ & Set of CT images \\
$\mathcal{M}$ & Set of PET images \\
$\mathcal{Z}$ & Zero-padding mask tensor\\
$\mathcal{F}$ & Feature tensor \\
$\mathcal{D}$ & CT-PET-Clinical multimodal Dataset \\
\bottomrule
\end{tabular}%
}
\end{table}

\section{Related Works}\label{II}

\subsection{Pre-trained Models for Extracting Image Features}

The evolution of pre-trained models has fundamentally transformed image feature extraction, particularly in domains with limited annotated data. Early work by Shin et al.~\cite{shin2016MedImg} leveraged convolutional neural networks (CNNs) pre-trained on large-scale datasets and enhanced them through data augmentation techniques. While these CNN-based approaches improved performance on small medical datasets, they often struggled with spatial representation challenges and orientation-specific tasks.  

The introduction of Transformer-based architectures by Vaswani et al.~\cite{vaswani2017attention} marked a significant shift in representation learning. Although their self-attention mechanism effectively captured long-range dependencies, its direct application to images was computationally prohibitive. To address this, localized attention strategies proposed by Parmar et al.~\cite{parmar2018image} constrained self-attention to smaller, more manageable regions, making Transformers more practical for image feature extraction.  

A major breakthrough came with the Vision Transformer (ViT) by Dosovitskiy et al.~\cite{dosovitskiy2020image}, which processed images as sequences of non-overlapping patches. This approach not only simplified image representation but also achieved state-of-the-art performance, provided that large-scale datasets were available for pre-training. However, its reliance on massive pre-training data remained a challenge for specialized fields like medical imaging, where data collection is inherently limited.

To mitigate data dependency issues, Oquab et al.~\cite{oquab2023dinov2} introduced DINOv2, a self-supervised learning framework that employs an automated preprocessing pipeline and a teacher-student training approach to address data limitations. By integrating key elements of DINO and iBOT, it enhances transfer learning for both classification and dense prediction tasks. However, its effectiveness in multimodal settings remains uncertain. Additionally, Perez-Garcia et al.~\cite{pérez2025DINO} demonstrated that DINOv2 generalizes well in medical imaging by developing RAD-DINO, a biomedical image encoder pretrained solely on imaging data. Their findings show that RAD-DINO matches or outperforms state-of-the-art models trained with text supervision across multiple benchmarks, underscoring the potential of DINOv2-based self-supervised learning for biomedical feature extraction.

From CNN-based methods to the latest Transformer-driven architectures, all approaches share a common principle: they transform an input image into a high-level representation that can be applied across various tasks. This transformation process has driven renewed interest in how large-scale pre-training, effective data augmentation, and self-supervised learning can enhance performance, even in data-scarce domains.  

\subsection{Multimodal Fusion Using Self-Attention}

Integrating heterogeneous data sources is a fundamental challenge in multimodal learning. Self-attention mechanisms have emerged as a powerful solution, offering a flexible approach to capturing both local and global relationships across modalities~\cite{baltruvsaitis2018multimodal}.  

One notable example is the Attention Feature Fusion framework by Dai et al.~\cite{dai2021attentional}, which refines multimodal representations through a multi-scale Channel Attention Module. This method dynamically adjusts the relative importance of each modality while preserving essential information via skip connections, demonstrating how self-attention can enhance task-relevant feature extraction.  

Originally developed for natural language processing, self-attention has naturally extended to vision and multimodal contexts. Multi-head attention, which divides its focus across multiple subspaces, enables parallelized attention computations, making models more expressive and better equipped to handle spatial, temporal, and semantic variations across data streams.  

By facilitating cross-modal interactions, self-attention has become a cornerstone of advanced multimodal fusion strategies. Its adaptability and ability to dynamically highlight relevant information from each data source make it an essential tool for improving the integration of diverse modalities.

\section{TENSOR NOTATIONS AND PROBLEM FORMULATION}\label{III}

\subsection{Tensor Notations}

Scalars, vectors, matrices, and tensors are denoted by \(x\), \(\mathbf{x}\), \(\mathbf{X}\), and \(\mathcal{X}\), respectively. A tensor generalizes vectors and matrices to higher dimensions, making it a suitable representation for high-dimensional data.

An order-\(d\) tensor is defined as \(\mathcal{T} \in \mathbb{R}^{I_1 \times I_2 \times \cdots \times I_d}\), where each dimension is referred to as a mode. The elements of the tensor are denoted by \(T_{i_1,i_2,\ldots,i_d}\), where \(i_1, i_2, \ldots, i_d\) represent coordinates in different dimensions.

\subsection{Diagnosis of Fever of Unknown Origin}

In the context of diagnosing FUO, the physician's decision-making process\cite{bharucha2017fuowithctpet}\cite{minamimoto2022fuowithctpet} can be formalized as follows. Let \(\mathcal{N} \in \mathbb{R}^{N \times n \times n}\) denote the set of CT images, \(\mathcal{M} \in \mathbb{R}^{M \times n \times n}\) denote the set of PET images, and \(\mathbf{a} \in \mathbb{R}^{a}\) represent the vector of relevant clinical features. By incorporating clinical expertise, the physician integrates these three inputs—\(\mathcal{N}\), \(\mathcal{M}\), and \(\mathbf{a}\)—within a tensor-based framework to derive a diagnostic result \(y\).

\paragraph{ROI Delineation}
To enable accurate localization of potential pathological regions, we first extract pertinent features from the CT and PET images. Denote the CT features as belonging to the feature space \(\mathcal{F}_{\text{CT}} \subset \mathbb{R}^{d_{\text{CT}}}\), and the PET features as belonging to the feature space \(\mathcal{F}_{\text{PET}} \subset \mathbb{R}^{d_{\text{PET}}}\), where \(d_{\text{CT}}\) and \(d_{\text{PET}}\) represent the dimensionalities of the respective feature spaces. By combining information from \(\mathcal{F}_{\text{CT}}\) and \(\mathcal{F}_{\text{PET}}\), a robust representation is obtained, capturing both structural (CT) and metabolic (PET) characteristics. This facilitates precise delineation of the region of interest (ROI) for subsequent analysis and therapy planning.

\paragraph{Diagnostic Decision}
Once the region of interest (ROI) is delineated, clinicians integrate clinical features \(\mathbf{a}\) to formulate the final diagnostic outcome \(y\). Formally, this process can be expressed as \(f_{clinician}\) in Eq.~\ref{clinican}:

\begin{equation}\label{clinican}
 y = f_{clinician}\bigl(\mathcal{F}_{\text{CT}}, \mathcal{F}_{\text{PET}}, \mathbf{a}\bigr)    
\end{equation}

where \(y\) represents the clinical decision function derived from evidence-based criteria and expert judgment. The resulting scalar \(y\) provides quantitative support for clinical decision-making, guiding targeted treatments and enabling close monitoring of disease progression.

\section{DATASET AND METHODS}\label{IV}

\subsection{Study populations and standard diagnostic work‑up}

The study adhered to the principles of the Declaration of Helsinki as revised in 2013\cite{goodyear2007declaration}. Ethical approval was granted by the Ethics Committee of West China Hospital, and the requirement for informed consent was waived due to the retrospective design of the study.

\begin{figure}[htbp]
  \centering
  \includegraphics[width=0.5\textwidth]{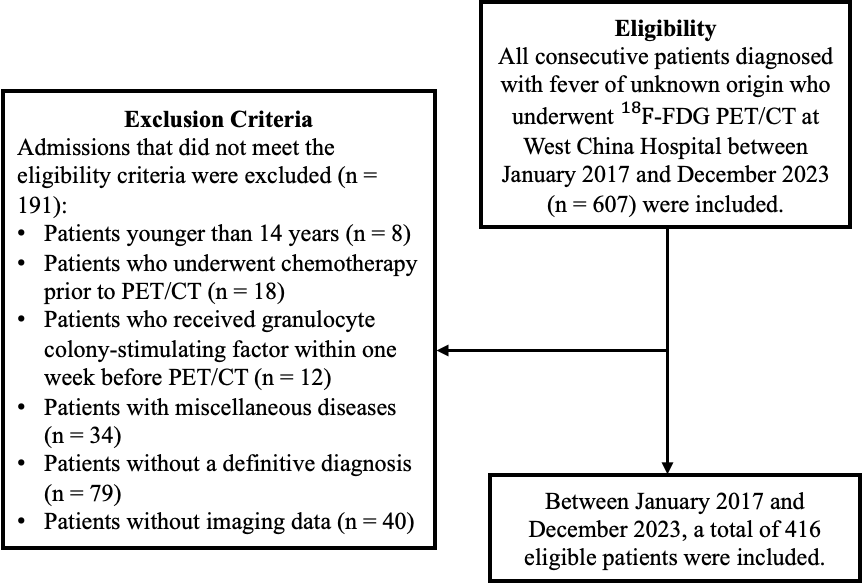}
  \caption{Flowchart of the inclusion and exclusion criteria.}
  \label{Flowchart of the inclusion and exclusion criteria}
\end{figure}

As illustrated in Fig.~\ref{Flowchart of the inclusion and exclusion criteria}, the medical records of patients older than 14 years with fever of unknown origin who underwent \textsuperscript{18}F-FDG PET/CT at the Nuclear Medicine Department of West China Hospital between January 2017 and December 2023 were retrospectively reviewed. Fever of unknown origin was defined by one of the following criteria: a duration exceeding three weeks, a temperature above 38.3℃ on at least three occasions, or an undetermined diagnosis despite comprehensive evaluations, including at least three outpatient visits or a minimum of three days of hospitalization\cite{muldersmanders2015fuo}. Patients without a definitive diagnosis and those with heterogeneous underlying conditions were excluded due to the limited sample size and the lack of common clinical characteristics. Ultimately, 416 patients with typical fever of unknown origin were included in the analysis, as detailed in Table~\ref{Patients}.

Standard diagnostic procedures included a comprehensive patient history, thorough physical examination, mandatory laboratory tests, and advanced imaging techniques such as contrast-enhanced computed tomography, magnetic resonance imaging, and nuclear medicine evaluations.

\begin{table}[t]
\centering
\caption{Patient Types and Disease Distribution.
In this table, we selected 416 patients and analyzed their patient characteristics and etiologies.}
\label{Patients}
\resizebox{\linewidth}{!}{%
\begin{tabular}{lcc}
\toprule
\textbf{Category} & \textbf{Number} & \textbf{Percentage (\%)} \\
\midrule
\multicolumn{3}{c}{\textbf{Patient Characteristics}} \\
\midrule
Age (47.07 \(\pm\) 17.45 [14-82]) & 416 & - \\
Male & 219 & 52.64 \\
Female & 197 & 47.36 \\
\midrule
\multicolumn{3}{c}{\textbf{Etiology Distribution}} \\
\midrule
\textbf{All Diseases } & 416 & - \\
\textbf{Benign Diseases } & 288 & 69.23 \\
\quad Immune-related & 140 & 33.65 \\
\quad \textbf{Infections } & 148 & 35.58 \\
\quad \quad Bacterial Infection & 103 & 24.76 \\
\quad \quad Viral Infection & 36 & 8.65 \\
\quad \quad Fungal Infection & 4 & 0.96 \\
\quad \quad Parasitic Infection & 5 & 1.20 \\
\textbf{Malignant Diseases }& 128 & 30.77 \\
\quad Solid Tumor & 110 & 26.44 \\
\quad Hematologic Malignancies & 18 & 4.33 \\
\bottomrule
\end{tabular}%
}
\end{table}

\subsection{Clinical Data Preparation}

A standard form was used to record clinical parameters such as age, sex, and laboratory features (blood count, C-reactive protein (CRP), serum ferritin (SF), procalcitonin (PCT), Alanine aminotransferase (ALT), Aspartate transaminase (AST), lactic dehydrogenase (LDH), Interleukin-2 Receptor (IL-2R), Interleukin-6 (IL-6), Interferon Gamma Release Assay (IGRA), Antinuclear Antibody (ANA), Anti-Neutrophil Cytoplasm Antibody (ANCA) etc.). Data on laboratory features were collected before the initial therapy and within 14 days before or after \textsuperscript{18}F-FDG PET/CT.

\textsuperscript{18}F-FDG was synthesized at the Nuclear Medicine Department of West China Hospital. All patients were needed to fast for 6 hours prior to intravenous administration of \textsuperscript{18}F-FDG, with a median activity of 5.18 MBq/Kg. We controlled the patient’s blood sugar levels to be below 7.0mmol/L. The CT scan parameters were 120 kV, 40 mAs, 5.0 mm slice thickness and 512 × 512 matrices. The PET scan parameters were 60 ± 5 min after tracer administration and 2.5 min per bed position. All patients underwent PET/CT scans with one of the following systems: Gemini GXL (Philips Corp, Netherlands), DISCOVERY 710 (GE, USA) and uMI780 (United Imaging, China). We used the acquired CT data to perform attenuation correction on all PET images.

\begin{figure}[t]
  \centering
  \includegraphics[width=0.5\textwidth]{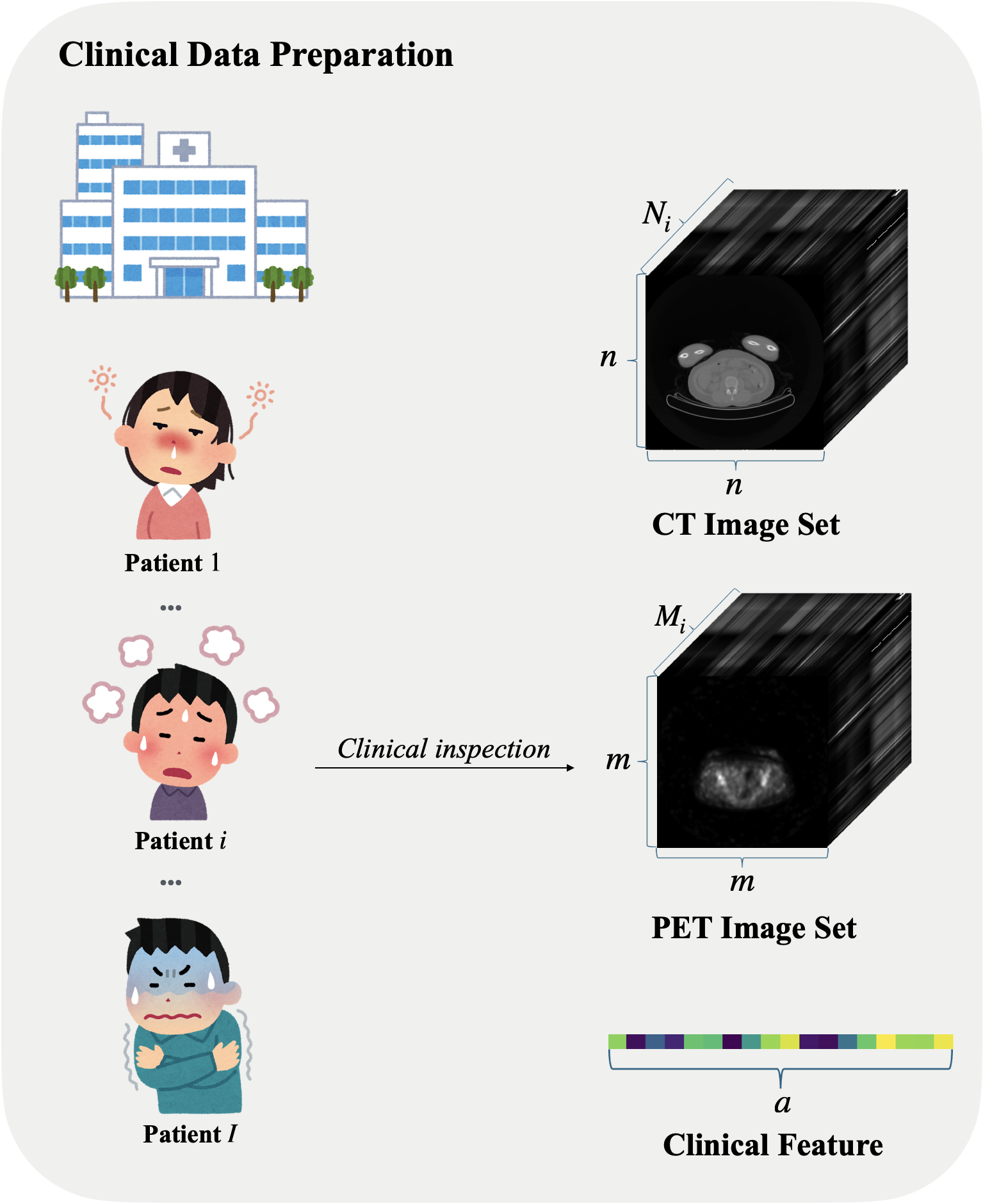}
  \caption{The progress of clinical data preparation. Each patient \(i\) is represented by a CT image set \(\mathcal{N}_i\), a PET image set \(\mathcal{M}_i\), a clinical feature vector \(\mathbf{a}_i\), and a one-hot label \(y_i\).}
  \label{Clinical Data Preparation}
\end{figure}

As illustrated in Fig.~\ref{Clinical Data Preparation}, our overall process is formally defined by representing each patient, indexed by \(i\), as an element of the dataset $\mathcal{D} = \{ (\mathcal{N}_i, \mathcal{M}_i, \mathbf{a}_i, y_i) \mid i = 1, \dots, I \}$. Specifically, for the \(i\)-th patient, the CT image set is denoted by \(\mathcal{N}_i \in \mathbb{R}^{N_i \times n \times n}\), comprising \(N_i\) slices, each with a resolution of \(n \times n\). Similarly, the PET image set is represented by \(\mathcal{M}_i \in \mathbb{R}^{M_i \times m \times m}\), containing \(M_i\) slices, each with a resolution of \(m \times m\). In addition, the clinical features are encapsulated in the vector \(\mathbf{a}_i \in \mathbb{R}^a\), which comprises \(a\) numerical attributes, and \(y_i\) denotes the one-hot encoded label. In total, the dataset comprises \(I\) patients.

\subsection{Methodology}

\subsubsection{Image Feature Extraction}

\begin{figure*}[htb]
    \centering
    \includegraphics[width=\textwidth]{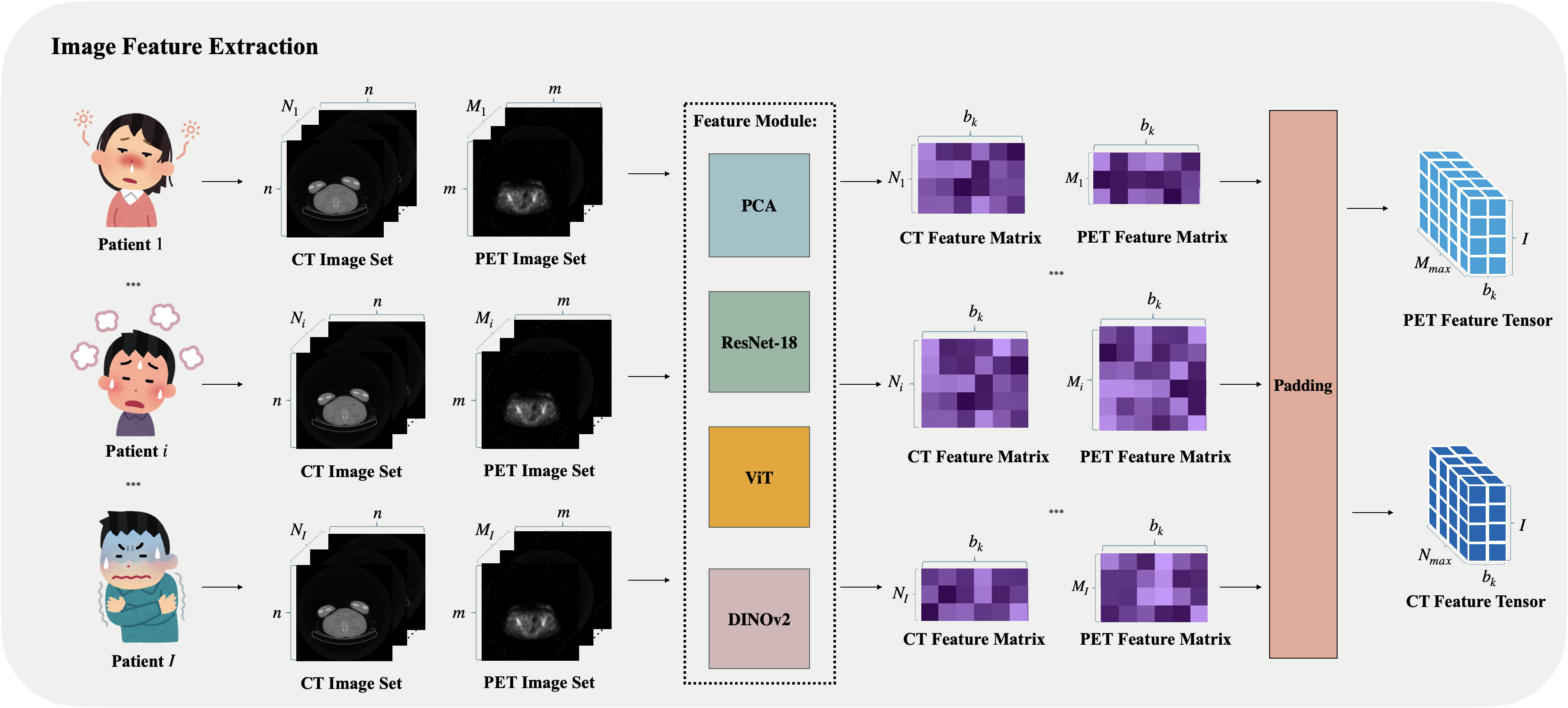}
    \caption{The progress of image feature extraction. For each model \(k\), slice features are extracted using PCA, ResNet-18, ViT, and DINOv2. For a patient \(i\) with \(N_i\) CT slices, the features are stacked into \(\mathbf{F}^{CT}_{ki}\). Zero-padding extends slices to \(N_{\max} = \max_i\{N_i\}\), forming \(\mathcal{F}_k^{\text{CT}}\). PET slices are processed similarly to yield \(\mathcal{F}_k^{\text{PET}}\).}
    \label{Image Feature Extraction}
\end{figure*}

FUO can be broadly categorized into infectious diseases, immune-related diseases, malignancies, and miscellaneous causes. \textsuperscript{18}F-FDG PET/CT provides both metabolic and anatomical insights, enabling more precise identification of FUO etiology. Infectious diseases such as pneumonia, soft tissue infections, endocarditis, and septicemia often exhibit focal or diffuse hypermetabolic activity on \textsuperscript{18}F-FDG PET/CT. Autoimmune and rheumatic diseases (e.g., large-vessel vasculitis, polymyalgia rheumatica, rheumatoid arthritis) typically show increased FDG uptake along inflamed vessels or joints. Malignancies, including solid tumors and myeloproliferative disorders, are likewise characterized by hypermetabolic lesions indicative of malignant activity.

To emulate radiologists’ diagnostic reasoning and extract relevant features from \textsuperscript{18}F-FDG PET/CT image sets, we employ an ensemble of state-of-the-art pretrained models uniquely suited for medical image analysis. Specifically, we integrate Google Research’s ViT \cite{dosovitskiy2020image}, torchvision’s ResNet-18 \cite{he2016deep}, and Meta’s self-supervised learning model DINOv2 \cite{oquab2023dinov2}, collectively forming a robust foundation for analyzing the intricate features present in \textsuperscript{18}F-FDG PET/CT data.

ResNet-18, with its efficient convolutional architecture, excels at capturing localized features critical for detecting hypermetabolic regions or structural abnormalities \cite{he2016deep}. ViT leverages patch-based tokenization and Transformer-driven encoding to capture global contextual information, facilitating the identification of complex metabolic and anatomical patterns across the entire image \cite{dosovitskiy2020image}. Meanwhile, DINOv2, trained on over 140 million images in a self-supervised manner, provides robust and generalizable feature representations, even in data-limited medical imaging scenarios \cite{oquab2023dinov2}.

As illustrated in Fig.~\ref{Image Feature Extraction}, image features from PET/CT scans are extracted using four distinct methods: PCA \cite{abdi2010pca}, ResNet-18, ViT, and DINOv2. Each method generates a slice-level feature vector of dimension $b_k$ through $f_{pretrained}$ for every image slice. In PCA, each slice is flattened, and principal component analysis is applied to reduce dimensionality. With ResNet-18, the slice is processed through the network, and the output from the global average pooling layer serves as the feature representation. In ViT, each slice is partitioned into patches and processed by a Vision Transformer to generate an embedding. DINOv2 employs a similar Transformer-based architecture but integrates a specialized self-supervised pretraining strategy to enhance feature learning.

Assume that \(k\) is the index of the selected pre-trained model. For each patient \(i\), the \(N_i\) slice-level CT feature vectors are stacked row-wise to form a patient-specific feature matrix \(\mathbf{F}^{CT}_{ki} \in \mathbb{R}^{N_i \times b_k}\). Since the number of CT slices \(N_i\) varies across patients, zero-padding is applied to standardize the row dimension, setting \(N_{\max} = \max_i \{N_i\}\). These padded matrices are then aggregated across all patients to construct the CT feature tensor \(\mathcal{F}_k^{\text{CT}} \in \mathbb{R}^{N_{\max} \times b_k \times I}\).

A similar process is applied to PET slices. For each patient \(i\) with \(M_i\) PET slices, zero-padding ensures a uniform row dimension \(M_{\max} = \max_i \{M_i\}\), resulting in the PET feature tensor \(\mathcal{F}_k^{\text{PET}} \in \mathbb{R}^{M_{\max} \times b_k \times I}\). This process is outlined in Algorithm~\ref{algo:Image_Feature_Extraction}. Since CT and PET feature extraction follow the same methodology, we use CT as the primary example in the subsequent discussion.

\paragraph{PCA}
\begin{figure}[t]
  \centering
  \includegraphics[width=0.5\textwidth]{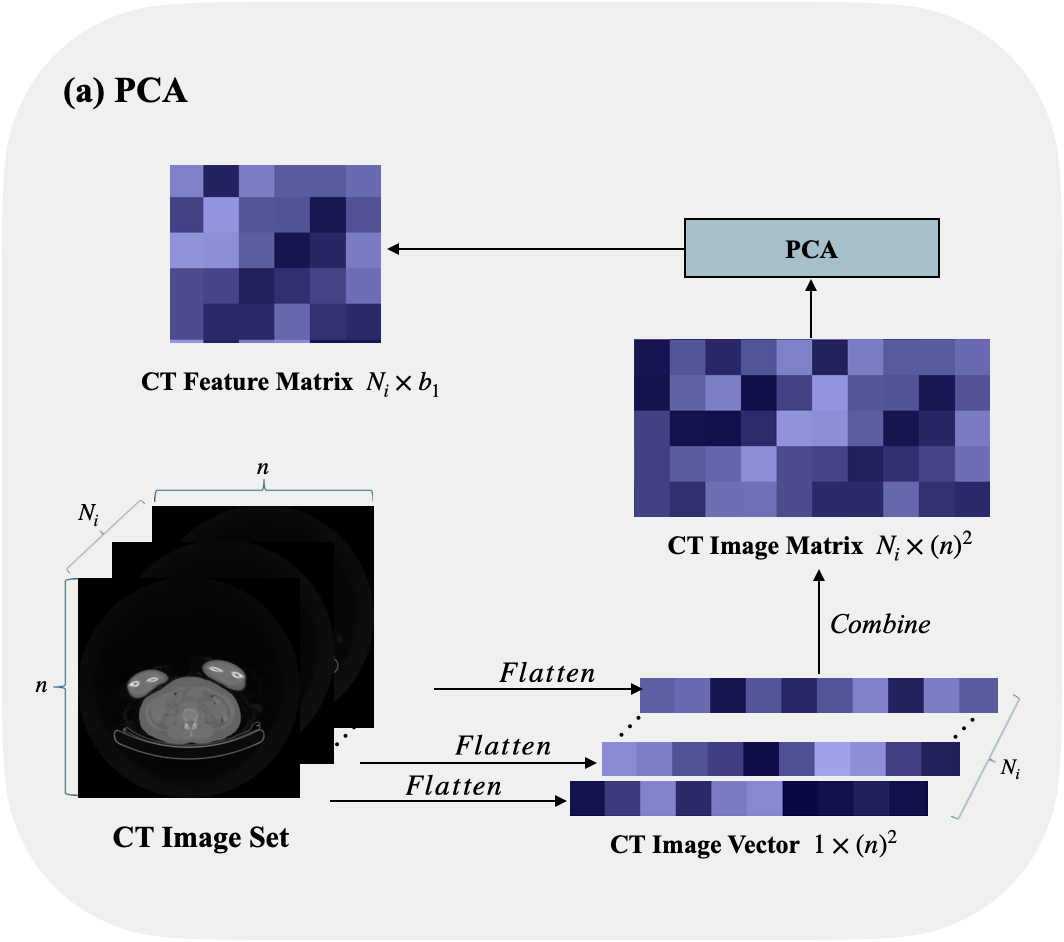}
  \caption{The progress of PCA based feature extraction. Each CT slice \(\mathbf{X}^{CT}\) is flattened into a vector and organized into a data matrix. After subtracting the mean at each pixel, PCA is applied via SVD to retain the top \(b_1\) eigenvectors, reducing the dimensionality of the data.}
  \label{PCA}
\end{figure}
As shown in Fig.~\ref{PCA}, Principal Component Analysis (PCA) serves as a baseline for comparison with advanced pre-trained models such as ResNet-18, ViT, and DINOv2. Each CT slice, denoted by \(\mathbf{X}^{CT}\), is treated as a single-channel \(n \times n\) square image. First, each image is flattened into a one-dimensional vector by sequentially concatenating its pixel values. These vectors are assembled into a data matrix, where each row represents a CT slice and each column corresponds to a specific pixel position. To correct for global intensity variations, the mean intensity at each pixel position across all slices is subtracted, yielding a zero-centered dataset. The covariance matrix is then computed to capture pixel intensity relationships. Singular value decomposition (SVD) is applied, and the top \(b_1\) eigenvectors corresponding to the largest eigenvalues are retained. These eigenvectors form a projection matrix that transforms the original high-dimensional data into a lower-dimensional space, reducing the number of features while preserving dominant variance for subsequent modeling tasks.

\paragraph{ResNet-18}
\begin{figure*}[t]
    \centering
    \includegraphics[width=\textwidth]{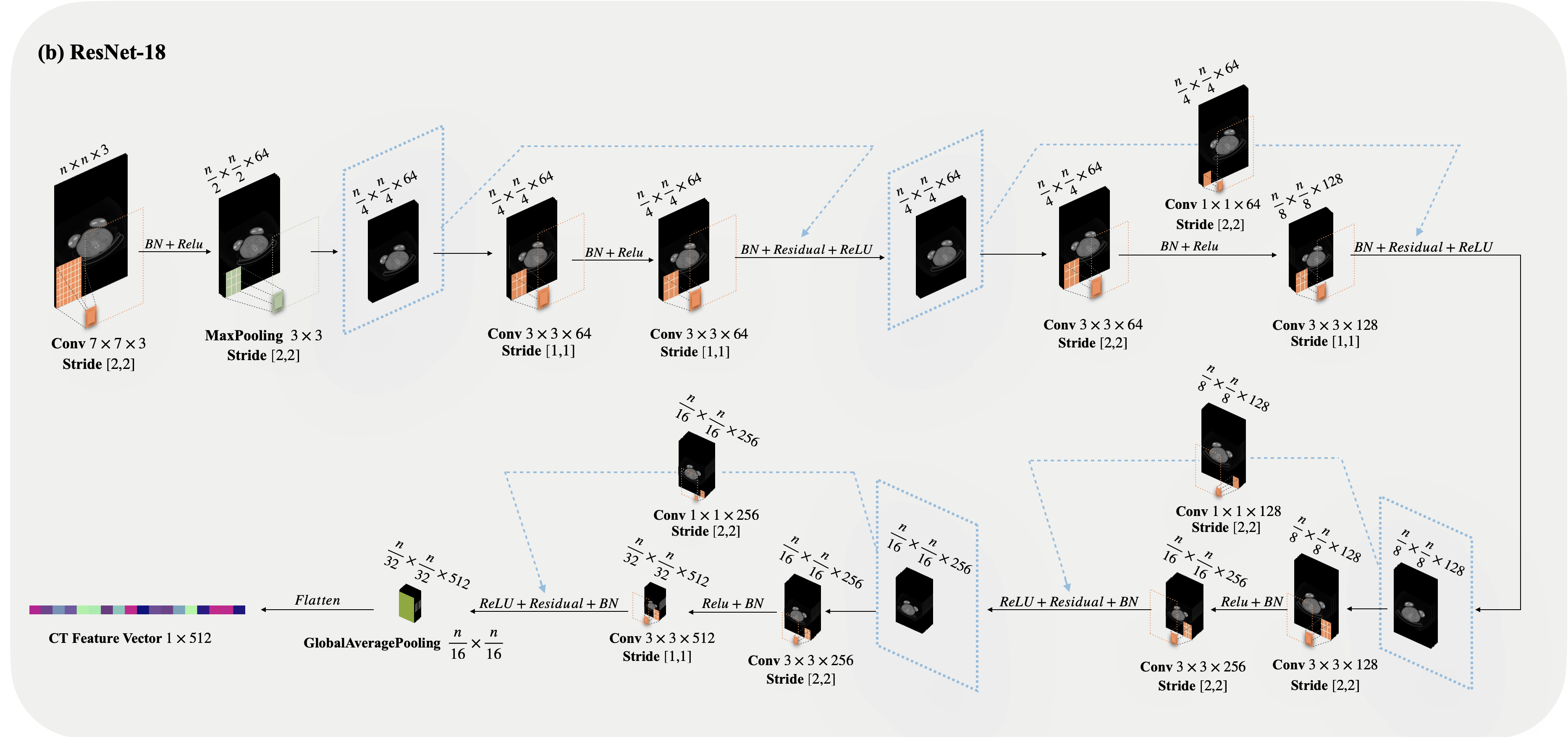}
    \caption{The progress of ResNet-18 based feature extraction. CT slices are replicated to create three channels and then processed through a \(7\times7\) convolution, max pooling, and residual blocks with skip connections. Global average pooling yields slice-level features \(f_{2i}\), which are stacked to form the patient-level feature matrix \(\mathbf{F}^{CT}_{2i}\).}
    \label{ResNet-18}
\end{figure*}

ResNet-18, a widely adopted convolutional architecture, employs hierarchical residual blocks to capture multi-scale feature representations. To accommodate the model's requirement for three-channel input, each CT slice \(\mathbf{X}^{CT}\) undergoes channel replication. As illustrated in Fig.~\ref{ResNet-18}, the architecture begins with a \(7 \times 7\) convolutional layer with stride 2, followed by max pooling for spatial downsampling. Subsequent residual blocks incorporate skip connections, enhancing gradient flow and improving training stability. We use a ResNet-18 variant pre-trained on ImageNet, enabling effective transfer learning. After the final residual block, global average pooling aggregates spatial information, producing compact slice-level features \(f_{2i} \in \mathbb{R}^{b_2}\), where \(b_2=512\). For patient \(i\), sequential aggregation of features across \(N_i\) CT slices forms the patient-level feature matrix \(\mathbf{F}^{CT}_{2i} \in \mathbb{R}^{N_i \times b_2}\), encapsulating volumetric information.

\paragraph{Vision Transformer}
\begin{figure}[t]
  \centering
  \includegraphics[width=0.5\textwidth]{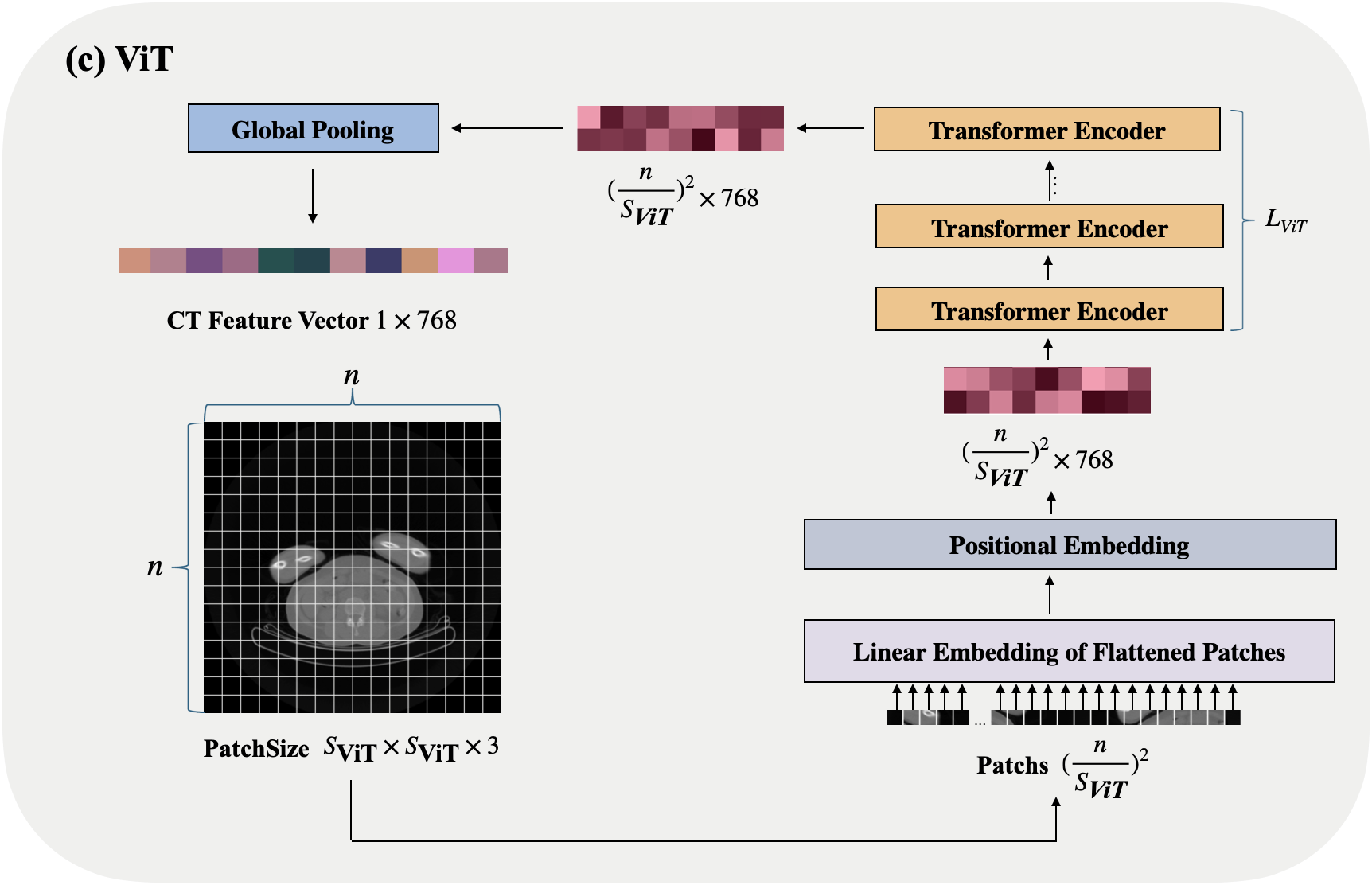}
  \caption{The progress of ViT based feature extraction. CT slices are replicated to three channels and split into non-overlapping patches. Each patch is embedded with positional encodings and processed through Transformer layers to form a slice-level feature vector \(f_{3i}\). Stacking these vectors yields the patient-level feature matrix \(\mathbf{F}^{CT}_{3i}\).}
  \label{ViT}
\end{figure}
We apply a pre-trained ViT without domain-specific fine-tuning. As shown in Fig.~\ref{ViT}, ViT processes each CT slice \(\mathbf{X}^{CT}\) by replicating it across three channels and partitioning it into non-overlapping patches of size \(S_{\text{ViT}} \times S_{\text{ViT}} \times 3\). Each flattened patch is projected into a high-dimensional embedding space, with positional encodings preserving spatial relationships. The sequence of embedded patches is passed through \(L_{ViT}\) Transformer Encoder layers, where Multi-Head Self-Attention captures both local and global dependencies, while skip connections stabilize training. The final token representations are aggregated into a single feature vector \(f_{3i} \in \mathbb{R}^{b_3}\), where \(b_3 = 768\). For each patient \(i\), stacking the slice-level feature vectors results in the patient-specific feature matrix \(\mathbf{F}^{CT}_{3i} \in \mathbb{R}^{N_i \times b_3}\).

\paragraph{DINOv2}
\begin{figure}[htbp]
  \centering
  \includegraphics[width=0.5\textwidth]{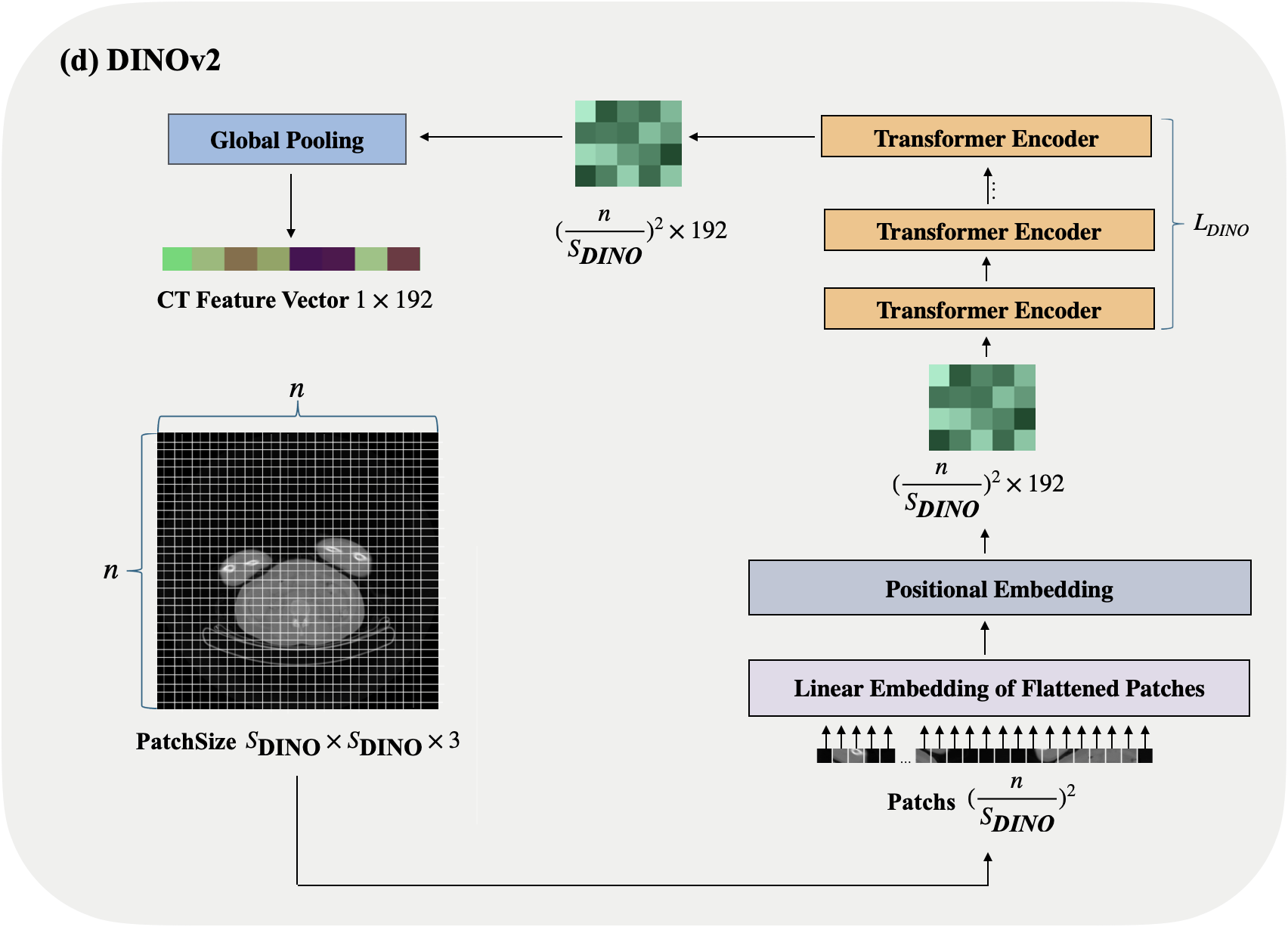}
  \caption{The progress of DINOv2 based feature extraction. CT slices are replicated to three channels and split into non-overlapping patches. Each patch is embedded with positional encodings and processed through Transformer layers to form a slice-level feature vector \(f_{3i}\). Stacking these vectors yields the patient-level feature matrix \(\mathbf{F}^{CT}_{4i}\).}
  \label{DINOv2}
\end{figure}
DINOv2 is a self-supervised learning framework enforcing multi-scale consistency through patch-based embeddings. As shown in Fig.~\ref{DINOv2}, it follows a similar patch extraction strategy as ViT, where each CT slice \(\mathbf{X}^{CT}\) is replicated across three channels and divided into non-overlapping patches of size \(S_{\text{DINO}} \times S_{\text{DINO}} \times 3\). To enhance feature robustness, multiple augmented views are generated using random cropping, color jittering, and geometric transformations. These views are processed by a student network and a teacher network, sharing the same backbone but differing in parameter update mechanisms. The teacher network, updated through a momentum-based approach, ensures stable training while guiding the student network to learn consistent and scale-invariant representations. In our pre-trained configuration, DINOv2 outputs a universal feature descriptor \(f_{4i} \in \mathbb{R}^{b_4}\), where \(b_4 = 192\). For each patient \(i\), stacking slice-level feature vectors produces the patient-specific feature matrix \(\mathbf{F}^{CT}_{4i} \in \mathbb{R}^{N_i \times b_4}\).

\begin{algorithm}[t]
\caption{Image Feature Extraction}\label{algo:Image_Feature_Extraction}
\begin{algorithmic}[1]
\STATE \textbf{Input:} 
\STATE \quad Dataset $\mathcal{D} = \{ (\mathcal{N}_i, \mathcal{M}_i, \mathbf{a}_i, y_i) \mid i = 1, \dots, I \}$;
\STATE \textbf{Multi-Modal Medical Image Feature Extraction}
    \FOR{$i = 1$ to $I$}
        \FOR{$j = 1$ to $N_{\max}$}
            \STATE Extract CT features using a pretrained model:
            \STATE $\mathbf{f}_{CT} \gets f_{\text{pretrained}}(\mathbf{X}_{CT})$;
            \STATE Aggregate CT feature matrix:
            \STATE $\mathbf{F}_{CT} \gets \mathbf{f}_{CT}$;
        \ENDFOR
        \STATE Pad and construct CT tensors:
        \STATE $\mathcal{F}_{CT} \gets \text{Pad}(\mathbf{F}_{CT})$;
        \FOR{$j = 1$ to $M_{\max}$}
            \STATE Extract PET features using a pretrained model:
            \STATE $\mathbf{f}_{PET} \gets f_{\text{pretrained}}(\mathbf{X}_{PET})$;
            \STATE Pad and aggregate PET features:
            \STATE $\mathbf{F}_{PET} \gets \mathbf{f}_{PET}$;
        \ENDFOR
        \STATE Pad and construct PET tensors:
        \STATE $\mathcal{F}_{PET} \gets \text{Pad}(\mathbf{F}_{CT})$;
        \STATE Construct the clinical feature matrix:
        \STATE $\mathbf{A} \gets \mathbf{a}$;
    \ENDFOR
\STATE \textbf{Output:} 
\STATE \quad CT feature tensor: $\mathcal{F}_{CT}$;
\STATE \quad PET feature tensor: $\mathcal{F}_{PET}$;
\STATE \quad Clinical feature matrix: $\mathbf{A}$;
\end{algorithmic}
\end{algorithm}

\subsubsection{Multimodal Fusion Classification Network}

\begin{figure*}[htb] 
    \centering
    \includegraphics[width=\textwidth]{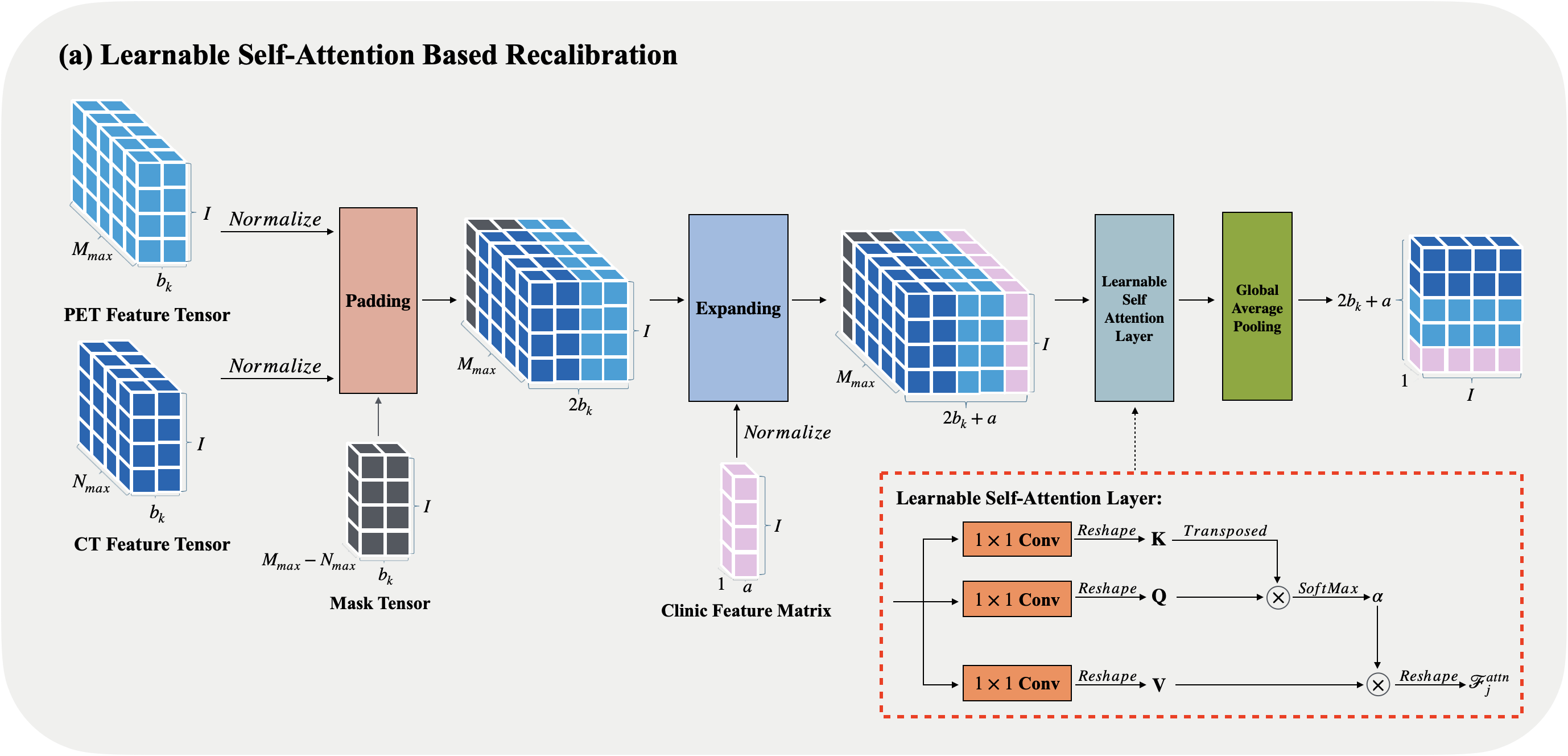}
    \caption{The architecture of Self-Attention-Based Feature Recalibration. CT, PET, and clinical features are fused into a unified tensor, followed by learnable self-attention to capture intra- and inter-modality dependencies. The recalibrated features are then pooled to generate the final representation.}
    \label{Self‐Attention‐Based Feature Recalibration}
\end{figure*}

\begin{figure}[h]
  \centering
  \includegraphics[width=0.5\textwidth]{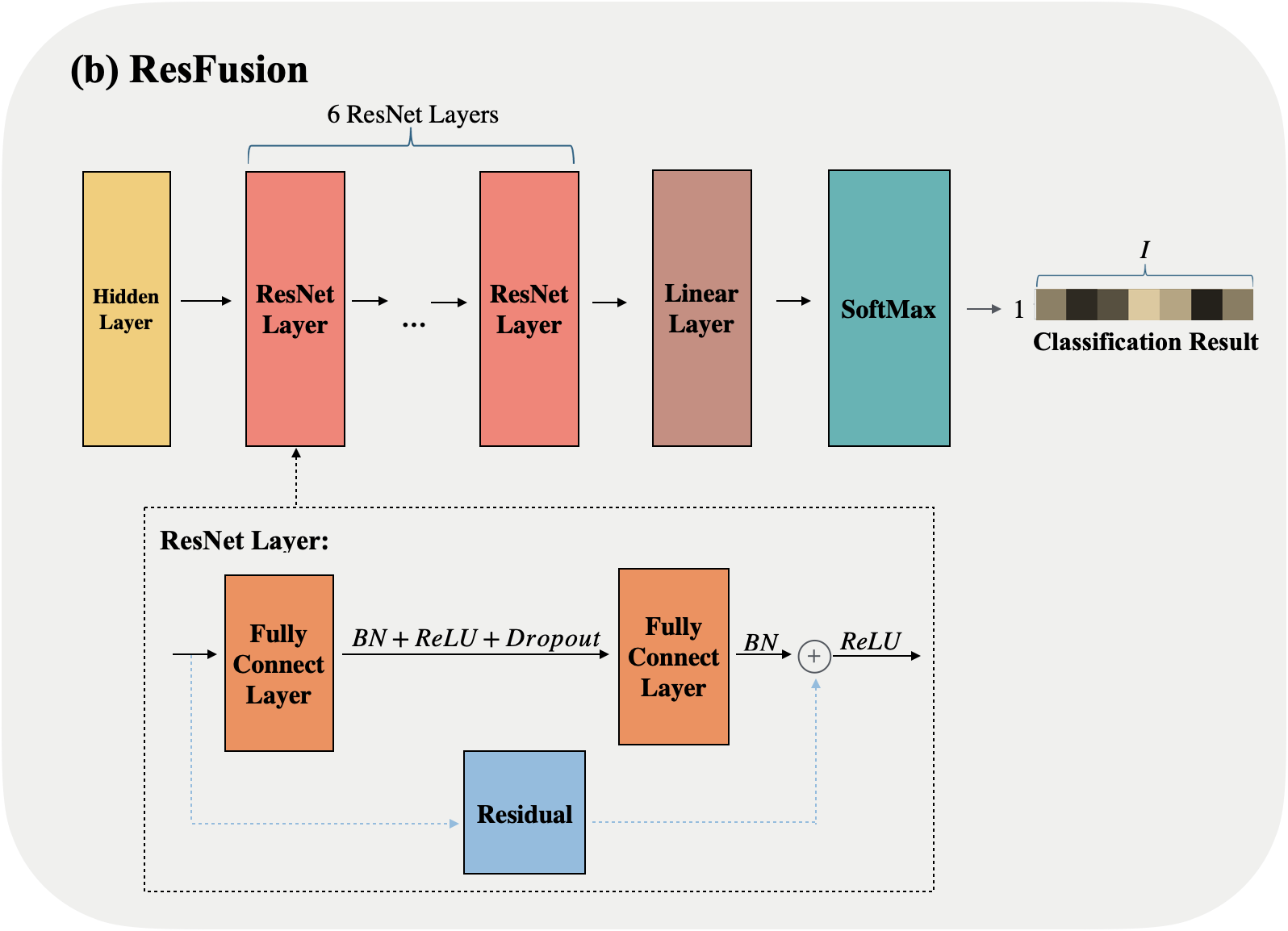}
  \caption{The architecture of the ResFusion model. Input features are processed through a hidden layer, followed by six ResNet layers with skip connections. A linear layer reduces the feature dimensions before classification is performed using SoftMax.}
  \label{ResFusion}
\end{figure}

A key component of our framework is the multimodal fusion classification network (MFCN), which comprises two main modules: data recalibration and classification.

\paragraph{Learnable Self-Attention Based Recalibration}
To emulate the diagnostic reasoning of clinical physicians, we introduce a learnable self-attention layer that dynamically captures intra- and inter-modality dependencies. The concatenated feature tensor is processed by the self-attention module, enabling the model to learn meaningful relationships among diverse data components. During backpropagation, this mechanism adjusts convolutional parameters, updating modality weights in a data-driven manner.

As illustrated in Fig.~\ref{Self‐Attention‐Based Feature Recalibration}, consider the CT feature tensor $\mathcal{F}^{CT}_k$ and the PET feature tensor $\mathcal{F}^{PET}_k$, and assume that $M_{\max} > N_{\max}$. First, we normalize the data along the patient dimension $I$. To ensure a consistent slice dimension across all patients, we apply a zero-padding mask tensor $\mathcal{Z} \in \mathbb{R}^{(M_{\max} - N_{\max}) \times b_k \times I}$, yielding the masked feature tensor $\mathcal{F}^{\text{masked}}_k \in \mathbb{R}^{M_{\max} \times 2b_k \times I}$.

Next, for each patient $i$, let $\mathbf{a}_i$ denote the clinical feature vector. By stacking these row-wise, we obtain the clinical feature matrix $\mathbf{F}^{\text{clinic}} \in \mathbb{R}^{a \times I}$. After normalizing across the patient dimension, we expand $\mathbf{F}^{\text{clinic}}$ into the tensor $\mathcal{F}^{\text{clinic}} \in \mathbb{R}^{M_{\max} \times a \times I}$. We then fuse $\mathcal{F}^{\text{masked}}_k$ with $\mathcal{F}^{\text{clinic}}$ to form the combined feature tensor $\mathcal{F}^{\text{fused}}_k \in \mathbb{R}^{M_{\max} \times (2b_k + a) \times I}$, thus ensuring uniform slice dimensions for all patient representations.

Given $\mathcal{F}^{\text{fused}}_k$, we employ three separate $1\times 1$ convolutional layers to compute the query, key, and value tensors, $\mathcal{Q}$, $\mathcal{K}$, and $\mathcal{V} \in \mathbb{R}^{M_{\max} \times (2b_k + a) \times I}$, respectively. These tensors are subsequently reshaped into matrices $Q,\; K,\; V \in \mathbb{R}^{I \times (M_{\max} \cdot (2b_k + a))}$.

We define the attention map $\boldsymbol{\alpha} \in \mathbb{R}^{(M_{\max} \cdot (2b_k + a)) \times (M_{\max} \cdot (2b_k + a))}$, where each element $\alpha_{ij}$ quantifies the similarity between the $i$th query and the $j$th key. This map is computed by multiplying $Q$ with the transpose of $K$, scaling by $\sqrt{I}$, and applying the softmax function. The resulting attention map is then used to weight the value matrix $V$, producing the output tensor $\mathbf{F}^{attn}_j$.

Finally, $\mathbf{F}^{attn}_j$ is reshaped back to its original spatial dimensions, yielding the recalibrated feature representation $\mathcal{F}^{attn}_k \in \mathbb{R}^{I \times M_{\max} \times (2b_k + a)}$.

This recalibration process enhances the model’s ability to focus on the most informative regions, thereby capturing fine-grained multimodal dependencies. Subsequently, global average pooling is applied along the $M_{\max}$ dimension to produce the fused matrix $\mathbf{F}^{fused}_k \in \mathbb{R}^{(2b_k + a) \times I}$. The detailed implementation of above is presented in Algorithm~\ref{algo:Learnable Self-Attention Based Recalibration}.

\paragraph{Multimodal Fusion and Classification}

Figure~\ref{ResFusion} illustrates the process. First, the input is passed through a hidden layer to learn an initial feature representation. Next, six consecutive ResNet layers are applied, each comprising two fully connected layers followed by batch normalization, ReLU activation, and dropout, with a skip connection adding the block input to its output.

After the ResNet layers, the aggregated feature map is processed by a linear layer for dimensionality reduction. Finally, a SoftMax function generates a probability distribution over the \(I\) target classes, yielding the final classification output. Algorithm~\ref{algo:ResFusion} details the implementation of the proposed ResFusion method.

\paragraph{Optimization}
The network is trained by minimizing the cross-entropy loss\cite{rumelhart1986crossenloss} between the model outputs and the ground-truth labels. Denote by \(\mathbf{p}_i\) the predicted probability vector for sample \(i\) and by \(\mathbf{y}_i\) its one-hot-encoded label. The loss over \(I\) samples is given by
\begin{equation}\label{loss}
\mathcal{L} = - \frac{1}{I} \sum_{i=1}^{I} \sum_{c=1}^{C} y_{i,c}\,\log\bigl(p_{i,c}\bigr),
\end{equation}
where \(C\) represents the number of target classes.

\begin{algorithm}[t]
\caption{Learnable Self-Attention Based Recalibration}\label{algo:Learnable Self-Attention Based Recalibration}
\begin{algorithmic}[1]
\STATE \textbf{Input:} 
\STATE \quad CT feature tensor: $\mathcal{F}_{CT}$;
\STATE \quad PET feature tensor: $\mathcal{F}_{PET}$;
\STATE \quad Clinical feature matrix: $\mathbf{A}$;
\STATE \quad Ground-truth label for each patient: $\mathbf{y}$;
\STATE \textbf{Initialization:} 
\STATE \quad Initialize convolution network parameters Conv$_{K}$, Conv$_{Q}$, Conv$_{V}$;
\STATE \textbf{Step 1: Tensor Alignment}
\STATE Normalize feature tensors:
\STATE \quad $\text{Norm}(\mathcal{F}^{CT})$, $\text{Norm}(\mathcal{F}^{PET})$;
\STATE Apply zero-padding for consistency:
\STATE \quad $\mathcal{F}^{masked} \gets \text{Pad}(\mathcal{F}^{CT}, \mathcal{F}^{PET}, \mathcal{Z})$;
\STATE Fuse normalized clinical and masked features:
\STATE \quad $\mathcal{F}^{fused} \gets \text{Concat}(\text{Expand}(\text{Norm}(\mathbf{A})), \mathcal{F}^{masked})$;
\STATE \textbf{Step 2: Learnable Self-Attention Based Recalibration}
\STATE Apply $1\times 1$ convolution and reshape the tensor:
\STATE \quad $\mathbf{K} \gets$ Reshape(Conv$_{K}$($\mathcal{F}^{fused}$));
\STATE \quad $\mathbf{Q} \gets$ Reshape(Conv$_{Q}$($\mathcal{F}^{fused}$));
\STATE \quad $\mathbf{V} \gets$ Reshape(Conv$_{V}$($\mathcal{F}^{fused}$));
\STATE Compute the attention map:
\STATE \quad $\mathbf{\alpha} \gets \text{SoftMax} \left( \frac{\mathbf{K}^T \mathbf{Q}}{\sqrt{I}} \right)$;
\STATE  Compute the attention tensor and reshape:
\STATE \quad $\mathbf{F}^{attn} \gets \text{Reshape}(\mathbf{V}\mathbf{\alpha})$;
\STATE Perform global average pooling along $M_{\max}$ dimension:
\STATE \quad $\mathbf{F}^{fused} \gets \text{AvgPool}(\mathcal{F}^{attn})$;
\STATE \textbf{Step 3: Optimization}
\STATE \quad Compute model loss via Eq.~(\ref{loss});
\STATE \quad Update Conv$_{K}$, Conv$_{Q}$, Conv$_{V}$ via gradient descent;
\STATE \textbf{Output:} Fused feature matrix $\mathbf{F}^{fuse}$.
\end{algorithmic}
\end{algorithm}

\begin{algorithm}[t]
\caption{ResFusion}\label{algo:ResFusion}
\begin{algorithmic}[1]
\STATE \textbf{Input:} 
\STATE \quad Fused feature matrix: $\mathbf{F}^{fused}$;
\STATE \quad Ground-truth label for each patient: $\mathbf{y}$;
\STATE \textbf{Initialization:} 
\STATE Initialize weight matrices: 
\STATE \quad $\mathcal{W} = { (\mathbf{W}_q^p \mid p = 1, \dots, 6; q=1,2), \mathbf{W}_h, \mathbf{W}_{linear})}$;
\STATE Initialize bias terms: 
\STATE \quad $\mathbf{b}={(b_q^p \mid p = 1, \dots, 6; q=1,2), b_h, b_{linear})}$;
\STATE \textbf{Step 1: ResFusion}
\STATE Processed through a hidden layer
\STATE \quad $\mathbf{H}_0 \gets\text{ReLU}(\text{BN}(\mathbf{F}^{fused}\mathbf{W}_h+b_h))$;
\STATE Feature processing via ResNet layers:
\FOR{$p=1$ to 6}
    \STATE $\mathbf{H}_p \gets \text{ReLU} \left( \text{BN} \left(\mathbf{H}_{p-1} \mathbf{W}_1^p + b_1^p \right) \right)$;
    \STATE $\mathbf{H}_p \gets \text{Dropout}(\mathbf{H}_p)$;
    \STATE $\mathbf{H}_p \gets \text{BN} \left(\mathbf{H}_p \mathbf{W}_2^p + b_2^p \right) + \mathbf{H}_{p-1}$;
    \STATE $\mathbf{H}_p \gets \text{ReLU}(\mathbf{H}_p)$ ;  
\ENDFOR
\STATE Classification via linear and SoftMax layers
\STATE \quad $\hat{\mathbf{y}}\gets\text{SoftMax}(\mathbf{H}_6\mathbf{W}_{linear}+b_{linear})$;
\STATE \textbf{Step 2: Optimization}
\STATE \quad Compute model loss via Eq.~(\ref{loss});
\STATE \quad Update $\mathcal{W}$, $\mathbf{b}$ via gradient descent;
\STATE \textbf{Output:} Predicted label for each patient $\hat{\mathbf{y}}$.
\end{algorithmic}
\end{algorithm}

\section{EXPERIMENTS}\label{V}

In this study, all experiments were conducted on a server running Ubuntu 24.04.1 LTS (Long-Term Support) and equipped with an Intel Core i5-13600KF processor (base frequency: 3.50 GHz, turbo frequency: 5.10 GHz), 32 GB of RAM, and a GeForce RTX 4090 GPU. Additionally, a high-performance computing server was utilized, featuring an Intel Xeon E5-2690 v4 CPU (2.60 GHz) with 56 cores, 377 GB of RAM, and two Tesla V100 GPUs (32 GB VRAM each). The MedMimic code is implemented based on PyTorch 2.5.1\cite{paszke2019pytorch}.

To thoroughly validate the framework's robustness and performance, the dataset was randomly shuffled and partitioned into five equal folds for 5-fold cross-validation\cite{kohavi19955cv}. In each iteration, one fold was used as the testing set while the remaining four folds served as the training set. The hyperparameter configuration that yielded the best performance on the training sets was selected as optimal. Specifically, the CNN hidden layer was set to a dimensionality of 256, the initial learning rate was set to 0.001 and gradually decayed using a cosine annealing scheduler\cite{loshchilov2016cas}, and the Adam optimizer\cite{kingma2014adam} was employed for its adaptive moment estimation capabilities.

\subsection{Compared Methods}

To systematically evaluate the feasibility of early diagnosis for FUO causes and assess the impact of incorporating patient characteristics, laboratory results, and \textsuperscript{18}F-FDG PET/CT imaging data, we categorized the implemented models into three groups, as detailed in Table~\ref{All Implemented Methods and Their Corresponding Input Data Forms}.

The first category includes baseline ML models: Logistic Regression\cite{cox1958regression}, Random Forest\cite{breiman2001rf}, Support Vector Machine\cite{vapnik2013svm}, and XGBoost\cite{chen2016xgboost}. These models were trained with various input configurations, including clinical data alone, CT alone, and PET alone. Despite their simplicity, they offer interpretable outputs and serve as robust benchmarks for comparison.

The second category consists of single-modality DL models, including convolutional neural network (CNN)\cite{lecun1989cnn}\cite{tajbakhsh2016cnnmed}, long short-term memory network (LSTM)\cite{hochreiter1997lstm}, and bidirectional LSTM(Bi-LSTM)\cite{graves2005bilstm}. These models are designed to extract complex patterns from a specific data type, such as clinical records, CT, or PET images. While leveraging deep architectures to uncover intricate relationships, they may overlook complementary information available across multiple modalities.

The third category is our MFCN model based on an attention mechanism, integrating clinical data, CT, and PET. By emphasizing the most informative features from each modality, this model enhances diagnostic performance, particularly in scenarios requiring heterogeneous data integration.

We systematically compare baseline ML models, single-modality DL models, and our fusion model to evaluate the impact of input modalities and fusion strategies on early FUO diagnosis.

\begin{table*}[ht]
\centering
\caption{All Implemented Methods and Their Corresponding Input Data Forms.}
\label{All Implemented Methods and Their Corresponding Input Data Forms}
\resizebox{\textwidth}{!}{%
\begin{tabular}{llccccccccc}
\toprule
\textbf{Category} & \textbf{Methods} & \textbf{Clinical Data} & \textbf{CT} & \textbf{PET} & \textbf{CT + PET} & \textbf{Clinical Data + CT} & \textbf{Clinical Data + PET} & \textbf{Clinical Data + CT + PET} \\ 
\midrule
\multirow{4}{*}{\textbf{Baseline ML Models}} 
& Logistic Regression 
& \Checkmark & \Checkmark & \Checkmark & \XSolidBrush & \XSolidBrush & \XSolidBrush & \XSolidBrush \\
& Random Forest       
& \Checkmark & \Checkmark & \Checkmark & \XSolidBrush & \XSolidBrush & \XSolidBrush & \XSolidBrush \\
& Support Vector Machine 
& \Checkmark & \Checkmark & \Checkmark & \XSolidBrush & \XSolidBrush & \XSolidBrush & \XSolidBrush \\
& XGBoost 
& \Checkmark & \Checkmark & \Checkmark & \XSolidBrush & \XSolidBrush & \XSolidBrush & \XSolidBrush \\
\midrule
\multirow{3}{*}{\textbf{\centering Single-Modality DL Models}} 
& CNN 
& \Checkmark & \Checkmark & \Checkmark & \XSolidBrush & \XSolidBrush & \XSolidBrush & \XSolidBrush\\
& LSTM 
& \Checkmark & \Checkmark & \Checkmark & \XSolidBrush & \XSolidBrush & \XSolidBrush & \XSolidBrush\\
& Bi-LSTM 
& \Checkmark & \Checkmark & \Checkmark & \XSolidBrush & \XSolidBrush & \XSolidBrush & \XSolidBrush\\
\midrule
\multirow{1}{*}{\textbf{Multimodal Models}} 
& MFCN 
& -- & -- & -- & \Checkmark & \Checkmark & \Checkmark & \Checkmark \\
\bottomrule
\end{tabular}%
}
\begin{flushleft}
\textbf{Note:} “\Checkmark” indicates that the corresponding input data is utilized by the method, while “\XSolidBrush” signifies that the input data is not used. The symbol “--” represents that it is not applicable.
\end{flushleft}
\end{table*}

\subsection{Evaluation Metrics}

Given the class imbalance in our dataset, we employed the area under the receiver operating characteristic curve (AUROC)\cite{zhang2014macroauc} as the primary evaluation metric due to its robustness against skewed class distributions.

For a multi-class classification problem with label set \(\mathcal{C} = \{c_1, c_2, \dots, c_{|\mathcal{C}|}\}\), we adopt a one-vs-rest strategy. Specifically, for each class \(c_i\), it is treated as the positive class while all other classes are considered negative, reducing the problem to a binary classification task. The AUROC for this "\(c_i\) vs. rest" scenario is denoted as \(\text{AUROC}(c_i)\). A standard definition of \(\text{AUROC}(c_i)\) is given by the integral under the ROC curve:
\begin{equation}\label{auroc}
    \text{AUROC}(c_i) \;=\; \int_0^1 \mathrm{TPR}_{c_i}\!\bigl(\mathrm{FPR}_{c_i}^{-1}(\beta)\bigr)\, \rm{d}\beta
\end{equation}
where \(\mathrm{TPR}_{c_i}\) (true positive rate) and \(\mathrm{FPR}_{c_i}\) (false positive rate) are computed by varying the decision threshold.

Finally, we compute the macro-averaged AUROC across all classes:
\begin{equation}
    \text{AUROC}_{\mathrm{macro}} = \frac{1}{|\mathcal{C}|} \sum_{c_i \in \mathcal{C}} \text{AUROC}(c_i)
\end{equation}

\subsection{Experimental Results and Discussions}

\paragraph{Experimental Results}

In this study, we evaluate the performance of the proposed MFCN method across seven distinct diagnostic tasks and compare it with other approaches. Specifically, Tables~\ref{Table Task1} through~\ref{Table Task7} present the performance metrics for the following classification tasks:

\begin{itemize}
    \item \textbf{Task 1:} Differential diagnosis of benign and malignant diseases.
    \item \textbf{Task 2:} Differential diagnosis of immune and non-immune diseases.
    \item \textbf{Task 3:} Differential diagnosis of infectious and non-infectious diseases.
    \item \textbf{Task 4:} Differential diagnosis of malignant, infectious, and immune diseases.
    \item \textbf{Task 5:} Differential diagnosis of solid tumors, hematologic malignancies, and non-malignant cases.
    \item \textbf{Task 6:} Differential diagnosis of bacterial, non-bacterial, and non-infectious diseases.
    \item \textbf{Task 7:} Differential diagnosis of viral, non-viral, and non-infectious diseases.
\end{itemize}

This comprehensive evaluation highlights the robustness and diagnostic accuracy of the proposed method across a diverse range of clinical scenarios, demonstrating its potential for improving differential diagnosis in medical practice.

Furthermore, we highlight the best-performing result for each modality in red. The optimal result for each specific task is emphasized in bold red. Additionally, we use underlining to denote the best result for each modality among the seven single-modality comparison methods.

\begin{table*}[t]
\centering
\caption{Task 1: differential diagnosis of benign and malignant diseases.}
\label{Table Task1}
\resizebox{\textwidth}{!}{%
\begin{tabular}{l c *{4}{c} *{4}{c} *{4}{c} *{4}{c}}
\toprule
\multicolumn{17}{c}{\textbf{Single-Modal Methods}} \\
\midrule
\multirow{2}{*}{\textbf{Methods}} & \multirow{2}{*}{\textbf{Clinical Data}} & \multicolumn{8}{c}{\textbf{CT}} & \multicolumn{8}{c}{\textbf{PET}}  \\
\cmidrule(lr){3-10} \cmidrule(lr){11-18}
 & & PCA && ResNet-18 && ViT && DINOv2 && PCA && ResNet-18 && ViT && DINOv2   \\
\midrule
Logistic Regression    & \underline{\textcolor{red}{0.8324}} &0.5680& &\underline{0.7047}& &\underline{0.7125}& &0.6783& &0.4995& &\underline{0.7238}& &0.6702& &0.5907&\\
Random Forest          & 0.8159 &0.5188& &0.5991& &0.5436& &0.5962& &0.4862& &0.4953& &0.5303& &\underline{\textcolor{red}{0.7343}}&\\
Support Vector Machine & 0.7800 &0.5044& &0.5000& &0.6236 & &0.5440& &0.4902& &0.6298& & 0.4305& &0.6208&\\
XGBoost                & 0.8166 &0.5809& &0.6437& &0.6379& &\underline{\textcolor{red}{0.7233}}& &0.5668& &0.6704& &\underline{0.6893}& &0.6958&\\
CNN                    & 0.6319 &0.4655& &0.5080& & 0.5696&  &0.5000& & 0.5000& & 0.6000& &0.6103& &0.6201&\\
LSTM                   & 0.6505 &0.6074& &0.4164& &0.3863& &0.5043& &\underline{0.6472}& & 0.5999 && 0.5748 &&0.6121&\\
Bi-LSTM                & 0.6492 &\underline{0.6160}&  &0.4011& &0.4689& & 0.5572& &0.5398& & 0.5521 &  & 0.5686&  &0.6398&\\
\midrule
\multicolumn{17}{c}{\textbf{Multi-Modal Methods}} \\
\midrule
\multirow{2}{*}{\textbf{Methods}} & \multicolumn{4}{c}{\textbf{CT + PET}} & \multicolumn{4}{c}{\textbf{Clinical Data + CT}} & \multicolumn{4}{c}{\textbf{Clinical Data + PET}} & \multicolumn{4}{c}{\textbf{Clinical Data + CT + PET}} \\
\cmidrule(lr){2-5} \cmidrule(lr){6-9} \cmidrule(lr){10-13} \cmidrule(lr){14-17}
 & PCA & ResNet-18 & ViT & DINOv2 & PCA & ResNet-18 & ViT & DINOv2 & PCA & ResNet-18 & ViT & DINOv2 & PCA & ResNet-18 & ViT & DINOv2 \\
\midrule
MFCN 
 & 0.6028 & \textcolor{red}{0.7865} & 0.7832 & 0.7553 
 & 0.7447 & 0.7838 & \textcolor{red}{0.8269} & 0.7918 
 & 0.7155 & 0.7739 & 0.7825 & \textcolor{red}{0.7984} 
 & 0.7400 & 0.8614 & 0.9058 & \textcolor{red}{\textbf{0.9178}} \\
\bottomrule
\end{tabular}%
}

\end{table*}

\begin{table*}[t]
\centering
\caption{Task2: differential diagnosis of immune and non-immune diseases.}
\label{Table Task2}
\resizebox{\textwidth}{!}{%
\begin{tabular}{l c *{4}{c} *{4}{c} *{4}{c} *{4}{c}}
\toprule
\multicolumn{17}{c}{\textbf{Single-Modal Methods}} \\
\midrule
\multirow{2}{*}{\textbf{Methods}} & \multirow{2}{*}{\textbf{Clinical Data}} & \multicolumn{8}{c}{\textbf{CT}} & \multicolumn{8}{c}{\textbf{PET}}  \\
\cmidrule(lr){3-10} \cmidrule(lr){11-18}
 & & PCA && ResNet-18 && ViT && DINOv2 && PCA && ResNet-18 && ViT && DINOv2   \\
\midrule
Logistic Regression    & \underline{\textcolor{red}{0.7595}} &0.5400& &0.6254& &\underline{0.6872}&  &\underline{\textcolor{red}{0.6944}}& &0.5024& &\underline{0.6231}& &\underline{\textcolor{red}{0.7249}}& &0.6711&\\
Random Forest          & 0.6987 &0.4718& &\underline{0.6476}&  &0.6091& &0.5671& &0.4718& &0.5278& &0.5731& &0.6469&\\
Support Vector Machine & 0.6448 &\underline{0.5964}& &0.6019&  &0.6465& &0.5539& &0.5072& &0.4823& &0.6110& &0.5646&\\
XGBoost                & 0.7032 &0.5725& &0.6278& &0.6673& &0.6453& &\underline{0.5827}& &0.5625& &0.6619& &\underline{0.7097}&\\
CNN                    & 0.7130 &0.4005& &0.5268& &0.5441& &0.5642& &0.5000&  &0.5445& &0.5672& &0.5756&\\
LSTM                   & 0.7340 &0.5466& &0.5357& &0.5313& &0.5434& &0.5395& &0.5080& &0.4483& &0.5043&\\
Bi-LSTM                & 0.7308 &0.4617&  &0.4605& &0.4212& &0.4617& &0.5746& &0.5303& &0.4974& &0.5572&\\
\midrule
\multicolumn{17}{c}{\textbf{Multi-Modal Methods}} \\
\midrule
\multirow{2}{*}{\textbf{Methods}} & \multicolumn{4}{c}{\textbf{CT + PET}} & \multicolumn{4}{c}{\textbf{Clinical Data + CT}} & \multicolumn{4}{c}{\textbf{Clinical Data + PET}} & \multicolumn{4}{c}{\textbf{Clinical Data + CT + PET}} \\
\cmidrule(lr){2-5} \cmidrule(lr){6-9} \cmidrule(lr){10-13} \cmidrule(lr){14-17}
 & PCA & ResNet-18 & ViT & DINOv2 & PCA & ResNet-18 & ViT & DINOv2 & PCA & ResNet-18 & ViT & DINOv2 & PCA & ResNet-18 & ViT & DINOv2 \\
\midrule
MFCN 
 & 0.6531 & 0.7066 & \textcolor{red}{0.7494} & 0.7455 
 & 0.7621 & 0.8495 & 0.8788 & \textcolor{red}{0.8897} 
 & 0.7244 & \textcolor{red}{0.9056} & 0.8859 & 0.8776 
 & 0.8147 & 0.8909 & 0.8986 & \textcolor{red}{\textbf{0.9210}} \\
\bottomrule
\end{tabular}%
}

\end{table*}

\begin{table*}[t]
\centering
\caption{Task3: differential diagnosis of infectious and non-infectious diseases.}
\label{Table Task3}
\resizebox{\textwidth}{!}{%
\begin{tabular}{l c *{4}{c} *{4}{c} *{4}{c} *{4}{c}}
\toprule
\multicolumn{17}{c}{\textbf{Single-Modal Methods}} \\
\midrule
\multirow{2}{*}{\textbf{Methods}} & \multirow{2}{*}{\textbf{Clinical Data}} & \multicolumn{8}{c}{\textbf{CT}} & \multicolumn{8}{c}{\textbf{PET}}  \\
\cmidrule(lr){3-10} \cmidrule(lr){11-18}
 & & PCA && ResNet-18 && ViT && DINOv2 && PCA && ResNet-18 && ViT && DINOv2   \\
\midrule
Logistic Regression    & 0.5069 &0.4800& &0.6618& &0.6347& &\underline{\textcolor{red}{0.6792}}& &\underline{0.5548}& &0.7398& &0.7063& &0.6574&\\
Random Forest          & 0.4663 &0.4796& &0.5727& &0.6254& &0.6633& &0.5207& &0.7011& &0.6048& &0.5561&\\
Support Vector Machine & 0.4874 &0.5016& &0.6542& &0.6213& &0.5417& &0.5291& &0.6663& &0.6840& &0.6878&\\
XGBoost                & 0.5522 &0.5501& &\underline{0.6735}& &\underline{0.6447}& &0.6225& &0.5107& &\underline{\textcolor{red}{0.7544}}& &\underline{0.7392}& &0.7268&\\
CNN                    & 0.5600 &0.5257& &0.5219& &0.5000& &0.5230& &0.5167& &0.7359& &0.6874& &\underline{0.7515}&\\
LSTM                   & 0.5438 &\underline{0.6290}& &0.5386& &0.5262& &0.5925& &0.3414& &0.5299& &0.5311& &0.5632&\\
Bi-LSTM                & \underline{\textcolor{red}{0.6250}} &0.5969&  &0.5019& &0.5546& &0.6269& &0.4759& &0.4821& &0.5348& &0.5222&\\
\midrule
\multicolumn{17}{c}{\textbf{Multi-Modal Methods}} \\
\midrule
\multirow{2}{*}{\textbf{Methods}} & \multicolumn{4}{c}{\textbf{CT + PET}} & \multicolumn{4}{c}{\textbf{Clinical Data + CT}} & \multicolumn{4}{c}{\textbf{Clinical Data + PET}} & \multicolumn{4}{c}{\textbf{Clinical Data + CT + PET}} \\
\cmidrule(lr){2-5} \cmidrule(lr){6-9} \cmidrule(lr){10-13} \cmidrule(lr){14-17}
 & PCA & ResNet-18 & ViT & DINOv2 & PCA & ResNet-18 & ViT & DINOv2 & PCA & ResNet-18 & ViT & DINOv2 & PCA & ResNet-18 & ViT & DINOv2 \\
\midrule
MFCN 
 & 0.6104 & 0.6661 & \textcolor{red}{0.7346} & 0.6870 
 & 0.6747 &0.7154  & 0.7530 & \textcolor{red}{0.7746} 
 & 0.6475 & 0.7019 & \textcolor{red}{0.7747} & 0.7531 
 & 0.7012 & 0.7472 & \textcolor{red}{\textbf{0.8654}} & 0.8556 \\
\bottomrule
\end{tabular}%
}

\end{table*}

\begin{table*}[t]
\centering
\caption{Task4: differential diagnosis of malignant, infectious, and immune diseases.}
\label{Table Task4}
\resizebox{\textwidth}{!}{%
\begin{tabular}{l c *{4}{c} *{4}{c} *{4}{c} *{4}{c}}
\toprule
\multicolumn{17}{c}{\textbf{Single-Modal Methods}} \\
\midrule
\multirow{2}{*}{\textbf{Methods}} & \multirow{2}{*}{\textbf{Clinical Data}} & \multicolumn{8}{c}{\textbf{CT}} & \multicolumn{8}{c}{\textbf{PET}}  \\
\cmidrule(lr){3-10} \cmidrule(lr){11-18}
 & & PCA && ResNet-18 && ViT && DINOv2 && PCA && ResNet-18 && ViT && DINOv2   \\
\midrule
Logistic Regression    & \underline{\textcolor{red}{0.7104}} &0.5216& &0.5884& &0.6340& &0.5802& &\underline{0.5589}& &\underline{0.6273}& &\underline{0.6449}& &0.5778&\\
Random Forest          & 0.6566 &0.5227& &0.5172& &0.5277& &0.6391& &0.4807& &0.5229& &0.5435& &0.6291&\\
Support Vector Machine & 0.6274 &0.5295& &\underline{0.6463}& &\underline{0.6400}& &\underline{\textcolor{red}{0.6689}}& &0.4796& &0.6115& &0.6282& &0.6434&\\
XGBoost                & 0.6022 &0.5061& &0.5385& &0.6324& &0.5861&  &0.4957&  &0.6211&  &0.6298& &0.6874&\\
CNN                    & 0.6988 &0.5050& &0.5049& &0.5089& &0.5179& &0.5185& &0.5620& &0.5179& &0.5729&\\
LSTM                   & 0.6493 &\underline{0.5570}& &0.4656& &0.4848& &0.5134& &0.5242& &0.4813& &0.5128& & \underline{\textcolor{red}{0.7284}}&\\
Bi-LSTM                & 0.5819 &0.4681& &0.4390& &0.4557& &0.5158& &0.5288 & &0.4936& &0.5060& &0.6507&\\
\midrule
\multicolumn{17}{c}{\textbf{Multi-Modal Methods}} \\
\midrule
\multirow{2}{*}{\textbf{Methods}} & \multicolumn{4}{c}{\textbf{CT + PET}} & \multicolumn{4}{c}{\textbf{Clinical Data + CT}} & \multicolumn{4}{c}{\textbf{Clinical Data + PET}} & \multicolumn{4}{c}{\textbf{Clinical Data + CT + PET}} \\
\cmidrule(lr){2-5} \cmidrule(lr){6-9} \cmidrule(lr){10-13} \cmidrule(lr){14-17}
 & PCA & ResNet-18 & ViT & DINOv2 & PCA & ResNet-18 & ViT & DINOv2 & PCA & ResNet-18 & ViT & DINOv2 & PCA & ResNet-18 & ViT & DINOv2 \\
\midrule
MFCN 
 & 0.5902 & \textcolor{red}{0.7243} & 0.7184 & 0.7045 
 & 0.7374 & 0.8707 & \textcolor{red}{0.9081} & 0.8955
 & 0.7066 & 0.8470 & 0.8979 & \textcolor{red}{0.9034} 
 & 0.7455 & 0.8951 & 0.9110 & \textcolor{red}{\textbf{0.9137}} \\
\bottomrule
\end{tabular}%
}

\end{table*}

\begin{table*}[t]
\centering
\caption{Task5: differential diagnosis of solid tumors, hematologic malignancies, and non-malignant cases.}
\label{Table Task5}
\resizebox{\textwidth}{!}{%
\begin{tabular}{l c *{4}{c} *{4}{c} *{4}{c} *{4}{c}}
\toprule
\multicolumn{17}{c}{\textbf{Single-Modal Methods}} \\
\midrule
\multirow{2}{*}{\textbf{Methods}} & \multirow{2}{*}{\textbf{Clinical Data}} & \multicolumn{8}{c}{\textbf{CT}} & \multicolumn{8}{c}{\textbf{PET}}  \\
\cmidrule(lr){3-10} \cmidrule(lr){11-18}
 & & PCA && ResNet-18 && ViT && DINOv2 && PCA && ResNet-18 && ViT && DINOv2   \\
\midrule
Logistic Regression    & 0.6735 &0.4882& &0.5314& &0.5767& &\underline{0.6718}& &0.4843& &0.6192& &0.6466& &\underline{0.6629}&\\
Random Forest          & 0.6092 &0.5617& &0.5933& &0.5325& &0.6391& &0.4790& &0.5731&  &0.5990& &0.6413&\\
Support Vector Machine & 0.5602 &0.5387& &0.4872& &0.5264& &0.5713& &0.4744& &0.5847&  &0.5802&  &0.5745&\\
XGBoost                & \underline{\textcolor{red}{0.7983}} &0.5272& &\underline{0.6876}& &0.5119& &0.4943& &0.5173& &0.6213&  &0.6358&  &0.6148&\\
CNN                    & 0.6141 &0.5000& &0.5583& &0.5000& &0.5764& &0.5125& &0.6063& &0.5406& &0.5787&\\
LSTM                   & 0.5944 &0.4843& &0.4562& &\underline{0.6781}& &0.6284&  &0.5431& &0.6625& &\underline{\textcolor{red}{0.8188}}& &0.6625&\\
Bi-LSTM                & 0.5462 &\underline{\textcolor{red}{0.7062}}&  &0.4656& &0.6188& &0.5878& &\underline{0.5656}& &\underline{0.7594}& &0.6656& &0.6365&\\
\midrule
\multicolumn{17}{c}{\textbf{Multi-Modal Methods}} \\
\midrule
\multirow{2}{*}{\textbf{Methods}} & \multicolumn{4}{c}{\textbf{CT + PET}} & \multicolumn{4}{c}{\textbf{Clinical Data + CT}} & \multicolumn{4}{c}{\textbf{Clinical Data + PET}} & \multicolumn{4}{c}{\textbf{Clinical Data + CT + PET}} \\
\cmidrule(lr){2-5} \cmidrule(lr){6-9} \cmidrule(lr){10-13} \cmidrule(lr){14-17}
 & PCA & ResNet-18 & ViT & DINOv2 & PCA & ResNet-18 & ViT & DINOv2 & PCA & ResNet-18 & ViT & DINOv2 & PCA & ResNet-18 & ViT & DINOv2 \\
\midrule
MFCN 
 & 0.6073 & \textcolor{red}{0.7563} & 0.7236 & 0.7227 
 & 0.6990 & 0.8135 & 0.8033 & \textcolor{red}{0.8250} 
 & 0.6579 & 0.8759 & 0.8649 & \textcolor{red}{0.8791} 
 & 0.7262 & 0.8780 & 0.9012 & \textcolor{red}{\textbf{0.9291}} \\
\bottomrule
\end{tabular}%
}

\end{table*}

\begin{table*}[t]
\centering
\caption{Task6: differential diagnosis of bacterial, non-bacterial, and non-infectious diseases.}
\label{Table Task6}
\resizebox{\textwidth}{!}{%
\begin{tabular}{l c *{4}{c} *{4}{c} *{4}{c} *{4}{c}}
\toprule
\multicolumn{17}{c}{\textbf{Single-Modal Methods}} \\
\midrule
\multirow{2}{*}{\textbf{Methods}} & \multirow{2}{*}{\textbf{Clinical Data}} & \multicolumn{8}{c}{\textbf{CT}} & \multicolumn{8}{c}{\textbf{PET}}  \\
\cmidrule(lr){3-10} \cmidrule(lr){11-18}
 & & PCA && ResNet-18 && ViT && DINOv2 && PCA && ResNet-18 && ViT && DINOv2   \\
\midrule
Logistic Regression    & 0.6440 &0.5287& &\underline{0.5799}& &0.5843& &0.6709& &0.5303& &0.6181& &0.5783& &\underline{\textcolor{red}{0.7148}}&\\
Random Forest          & 0.5772 &0.4839& &0.5729& &0.5492& &0.6331& &0.5080& &0.5742& &0.5492&  &0.6549&\\
Support Vector Machine & 0.6028 &0.5085& &0.5051& &0.6754& &0.4585& &0.4786& &0.5810& &0.5263&  &0.5712&\\
XGBoost                & 0.5103 &0.4834& &0.5590& &0.5651& &0.6138& &0.4680& &\underline{0.6437}& &0.6810&  &0.6206&\\
CNN                    & \underline{\textcolor{red}{0.8188}} &0.4145& &0.5525& &0.5009& &0.5014& &0.3474& &0.5677& &0.5276& &0.5569&\\
LSTM                   & 0.6365 &\underline{0.5439}& &0.5539& &\underline{0.6780}& &\underline{\textcolor{red}{0.7234}}& &0.5431& &0.5318& &\underline{0.6951}& &0.6625&\\
Bi-LSTM                & 0.5947 &0.4537& &0.4537& &0.6188& &0.5987& &\underline{0.6365}& &0.4367& &0.6235& &0.5278&\\
\midrule
\multicolumn{17}{c}{\textbf{Multi-Modal Methods}} \\
\midrule
\multirow{2}{*}{\textbf{Methods}} & \multicolumn{4}{c}{\textbf{CT + PET}} & \multicolumn{4}{c}{\textbf{Clinical Data + CT}} & \multicolumn{4}{c}{\textbf{Clinical Data + PET}} & \multicolumn{4}{c}{\textbf{Clinical Data + CT + PET}} \\
\cmidrule(lr){2-5} \cmidrule(lr){6-9} \cmidrule(lr){10-13} \cmidrule(lr){14-17}
 & PCA & ResNet-18 & ViT & DINOv2 & PCA & ResNet-18 & ViT & DINOv2 & PCA & ResNet-18 & ViT & DINOv2 & PCA & ResNet-18 & ViT & DINOv2 \\
\midrule
MFCN 
 & 0.6093 & 0.7052 & 0.7240 & \textcolor{red}{0.7317} 
 & 0.5725 & \textcolor{red}{0.7450} & 0.7291 & 0.7308 
 & 0.6307 & \textcolor{red}{0.7890} & 0.7457 & 0.7546 
 & 0.6751 & 0.8792 & 0.8342 & \textcolor{red}{\textbf{0.8899}} \\
\bottomrule
\end{tabular}%
}

\end{table*}

\begin{table*}[t]
\centering
\caption{Task7: differential diagnosis of viral, non-viral, and non-infectious diseases.}
\label{Table Task7}
\resizebox{\textwidth}{!}{%
\begin{tabular}{l c *{4}{c} *{4}{c} *{4}{c} *{4}{c}}
\toprule
\multicolumn{17}{c}{\textbf{Single-Modal Methods}} \\
\midrule
\multirow{2}{*}{\textbf{Methods}} & \multirow{2}{*}{\textbf{Clinical Data}} & \multicolumn{8}{c}{\textbf{CT}} & \multicolumn{8}{c}{\textbf{PET}}  \\
\cmidrule(lr){3-10} \cmidrule(lr){11-18}
 & & PCA && ResNet-18 && ViT && DINOv2 && PCA && ResNet-18 && ViT && DINOv2   \\
\midrule
Logistic Regression    & \underline{\textcolor{red}{0.6527}} &0.4938& &0.6183& &\underline{0.6707}& &\underline{\textcolor{red}{0.6821}}& &0.5385& &0.6168& &0.6302&  &0.6552&\\
Random Forest          & 0.6354 &0.4822& &0.5582& &0.6566& &0.6590& &0.5088& &0.5392& &0.5615&  &\underline{\textcolor{red}{0.6733}}&\\
Support Vector Machine & 0.5496 &0.5209& &0.6394& &0.6141&  &0.5732& &0.4459& &0.5836& &0.5732&  &0.5441&\\
XGBoost                & 0.5858 &0.5359& &\underline{0.6421}& &0.5907&  &0.6146& &0.4943& &\underline{0.6237}& &\underline{0.6598}& &0.6246&\\
CNN                    & 0.6018 &0.5000& &0.5000& &0.5789& &0.6292& &0.4339& &0.5130& &0.5724& &0.5073&\\
LSTM                   & 0.5758 &\underline{0.6147}& &0.5620& &0.5433& &0.5257& &\underline{0.5680}& &0.5170& &0.5329& &0.5444&\\
Bi-LSTM                & 0.6039 &0.5850& &0.4630& &0.5133& &0.6350&  &0.5278& &0.5206& &0.6273& &0.5278&\\
\midrule
\multicolumn{17}{c}{\textbf{Multi-Modal Methods}} \\
\midrule
\multirow{2}{*}{\textbf{Methods}} & \multicolumn{4}{c}{\textbf{CT + PET}} & \multicolumn{4}{c}{\textbf{Clinical Data + CT}} & \multicolumn{4}{c}{\textbf{Clinical Data + PET}} & \multicolumn{4}{c}{\textbf{Clinical Data + CT + PET}} \\
\cmidrule(lr){2-5} \cmidrule(lr){6-9} \cmidrule(lr){10-13} \cmidrule(lr){14-17}
 & PCA & ResNet-18 & ViT & DINOv2 & PCA & ResNet-18 & ViT & DINOv2 & PCA & ResNet-18 & ViT & DINOv2 & PCA & ResNet-18 & ViT & DINOv2 \\
\midrule
MFCN 
 & 0.5984 & 0.7298 & 0.7078 & \textcolor{red}{0.7595} 
 & 0.6538 & 0.8362 & 0.8425 & \textcolor{red}{0.8626} 
 & 0.6378 & 0.7531 & \textcolor{red}{0.8904} & 0.8247 
 & 0.6752 & 0.8667 & 0.8688 & \textcolor{red}{\textbf{0.8969}} \\
\bottomrule
\end{tabular}%
}

\end{table*}

\begin{figure}[t]
  \centering
  \includegraphics[width=0.5\textwidth]{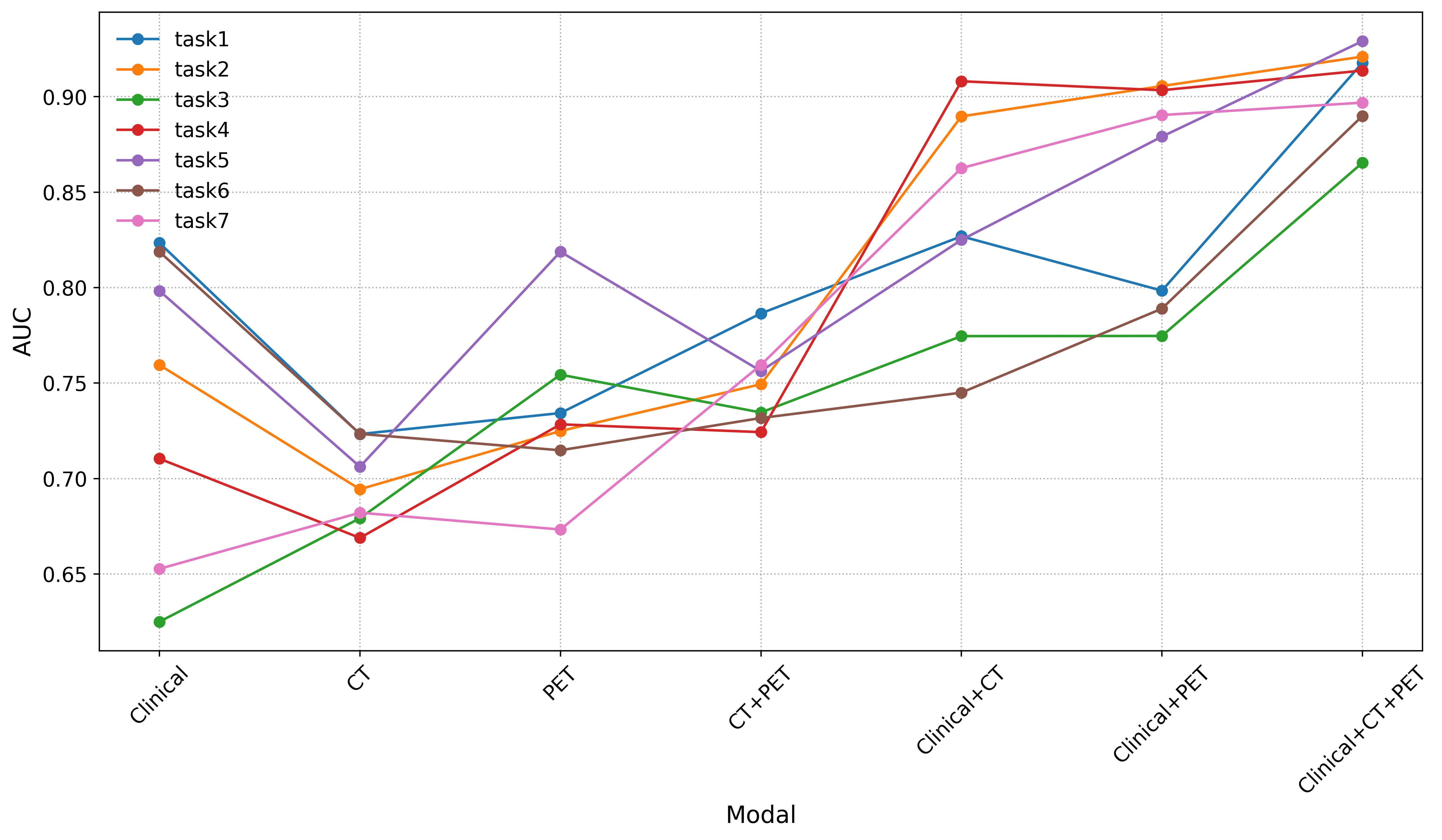}
  \caption{Line chart of the highest result for each modality across Tasks 1 to 7, showing that the "Clinical Data + CT + PET" configuration consistently achieved the best performance.}
  \label{best result}
\end{figure}

\begin{figure}[t]
  \centering
  \includegraphics[width=0.3\textwidth]{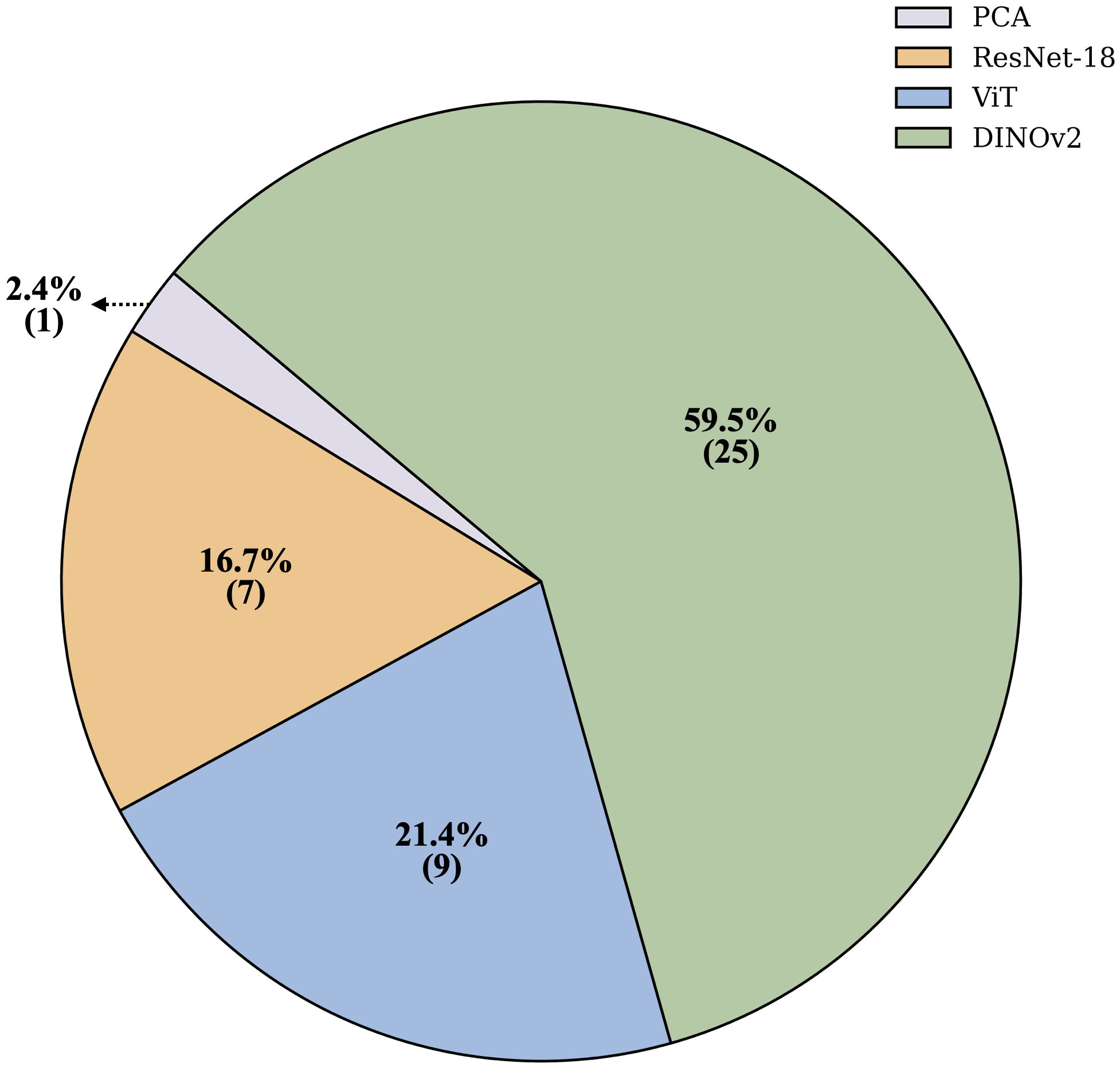}
  \caption{Performance comparison of pre-trained models for CT and PET feature extraction. DINOv2 achieved the best results in most cases, highlighting the effectiveness of Transformer-based models in medical imaging.}
  \label{pre-trained model}
\end{figure}

\begin{figure}[t]
  \centering
  \includegraphics[width=0.3\textwidth]{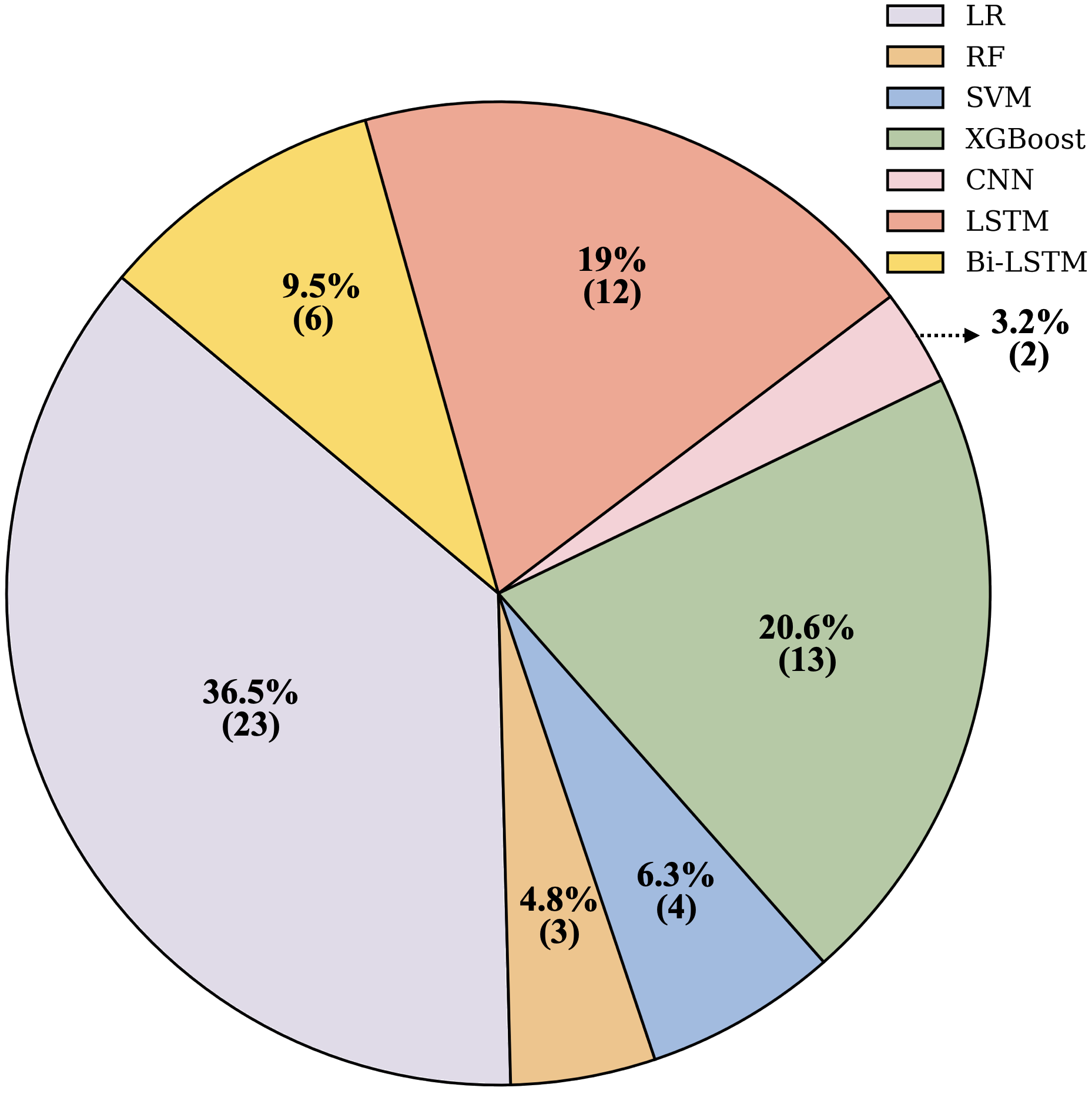}
  \caption{Optimal results of traditional classification methods in single-modal scenarios.}
  \label{single modal}
\end{figure}

\paragraph{Discussions}
From Task 1 to Task 7, the highest performance in each task was consistently achieved using the "Clinical Data + CT + PET" configuration, with accuracy ranging from 0.8654 to 0.9291. Furthermore, as illustrated in Figure~\ref{best result}, we have plotted a line chart highlighting the highest accuracy values for each modality across all tasks. These results demonstrate that our proposed MedMimic framework provides more accurate diagnostic assessments compared to single-modality or dual-modality approaches.

Additionally, as shown in Figure~\ref{pre-trained model}, we evaluated the effectiveness of various pre-trained models for feature extraction in each task using CT and PET modalities. Among the four methods, DINOv2 demonstrated a clear advantage, contributing to 59.5\% of the best-performing cases. ViT accounted for 21.4\%, ResNet-18 for 16.7\%, and PCA for only 2.4\%. Notably, Transformer-based models (DINOv2 and ViT) achieved superior performance in 79.9\% of cases, highlighting their enhanced capability in global information modeling and superior transferability to medical imaging compared to ResNet. Moreover, with PCA yielding the best performance in only a single instance, it is evident that extracting semantic features from medical images using pre-trained large models is significantly more effective than dimensionality reduction via PCA. This finding further supports the superior generalization capability of large models in medical imaging applications.

Finally, Figure~\ref{single modal} presents a summary of the optimal values obtained using traditional classification methods in single-modal scenarios. The experimental results indicate that logistic regression and XGBoost consistently achieved the best performance among these methods. However, their optimal values still exhibit a considerable performance gap compared to our proposed multimodal approach, further demonstrating the advantage of incorporating multi-source information for enhanced diagnostic accuracy.

\subsection{Ablation Studies}

To evaluate the contribution of different components within the MFCN model, we designed and tested seven experimental configurations: (1) without residual connections, (2) without attention mechanisms, (3) without dropout, (4) without convolutional layers, and (5) variations in the ResNet architecture with 3, 4, 5, or 6 layers. To ensure optimal feature representation, we utilized the DINOv2 model for semantic feature extraction from CT/PET images and conducted experiments using three data modalities: Clinical Data, CT, and PET.

\begin{figure}[htbp]
  \centering
  \includegraphics[width=0.5\textwidth]{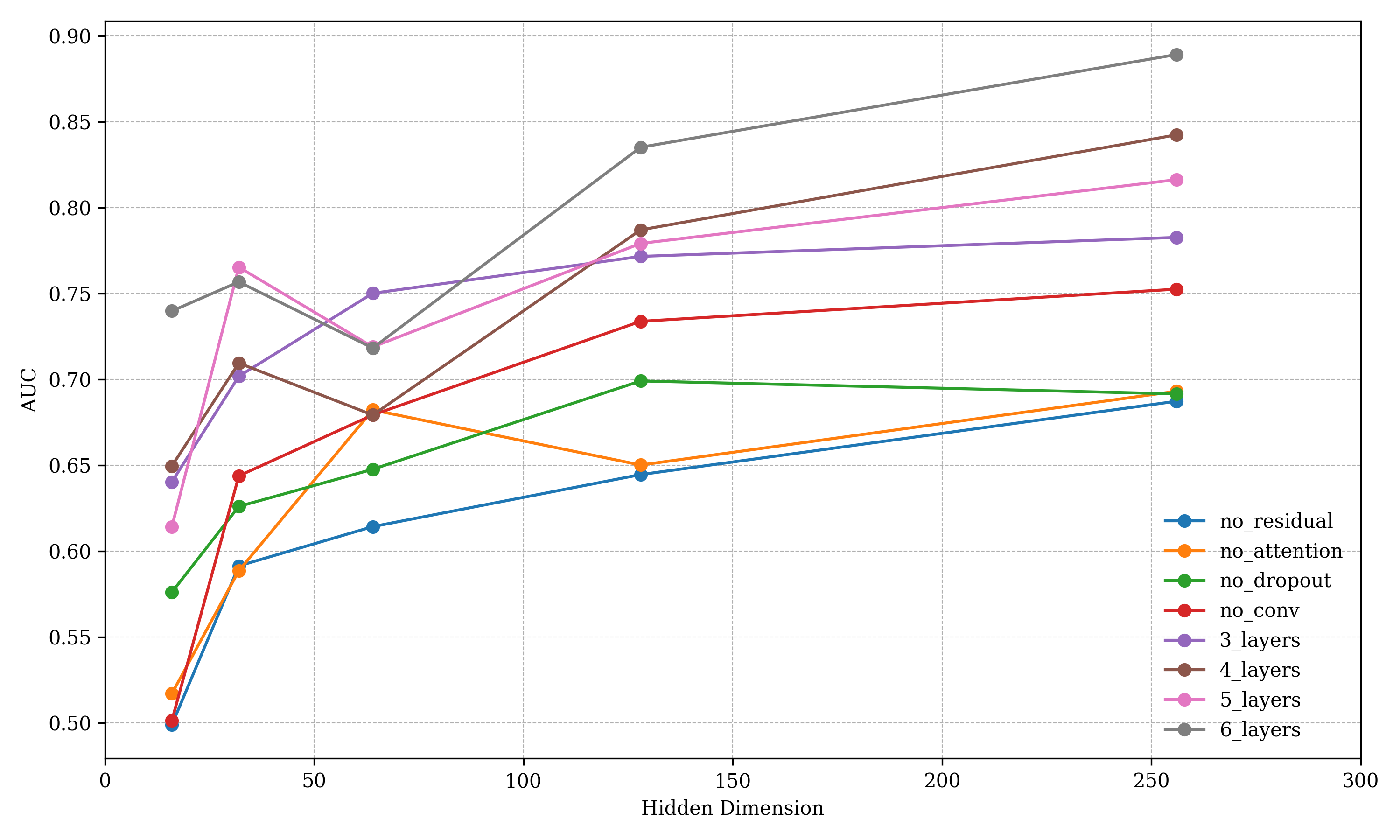}
  \caption{Comparison of seven experiments with different hidden dimensions in task 1.}
  \label{task1}
\end{figure}

\begin{figure}[htbp]
  \centering
  \includegraphics[width=0.5\textwidth]{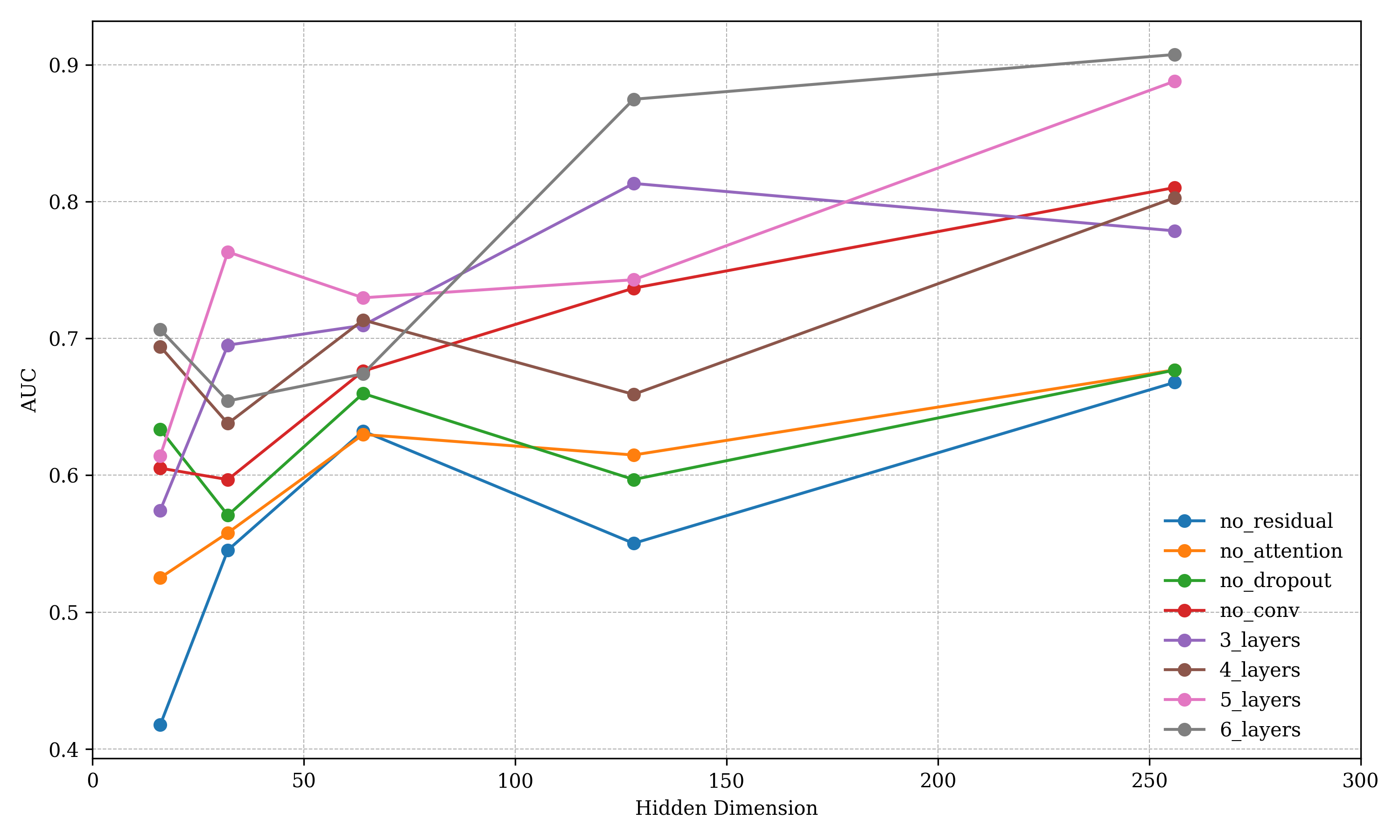}
  \caption{Comparison of seven experiments with different hidden dimensions in task 2.}
  \label{task2}
\end{figure}

\begin{figure}[htbp]
  \centering
  \includegraphics[width=0.5\textwidth]{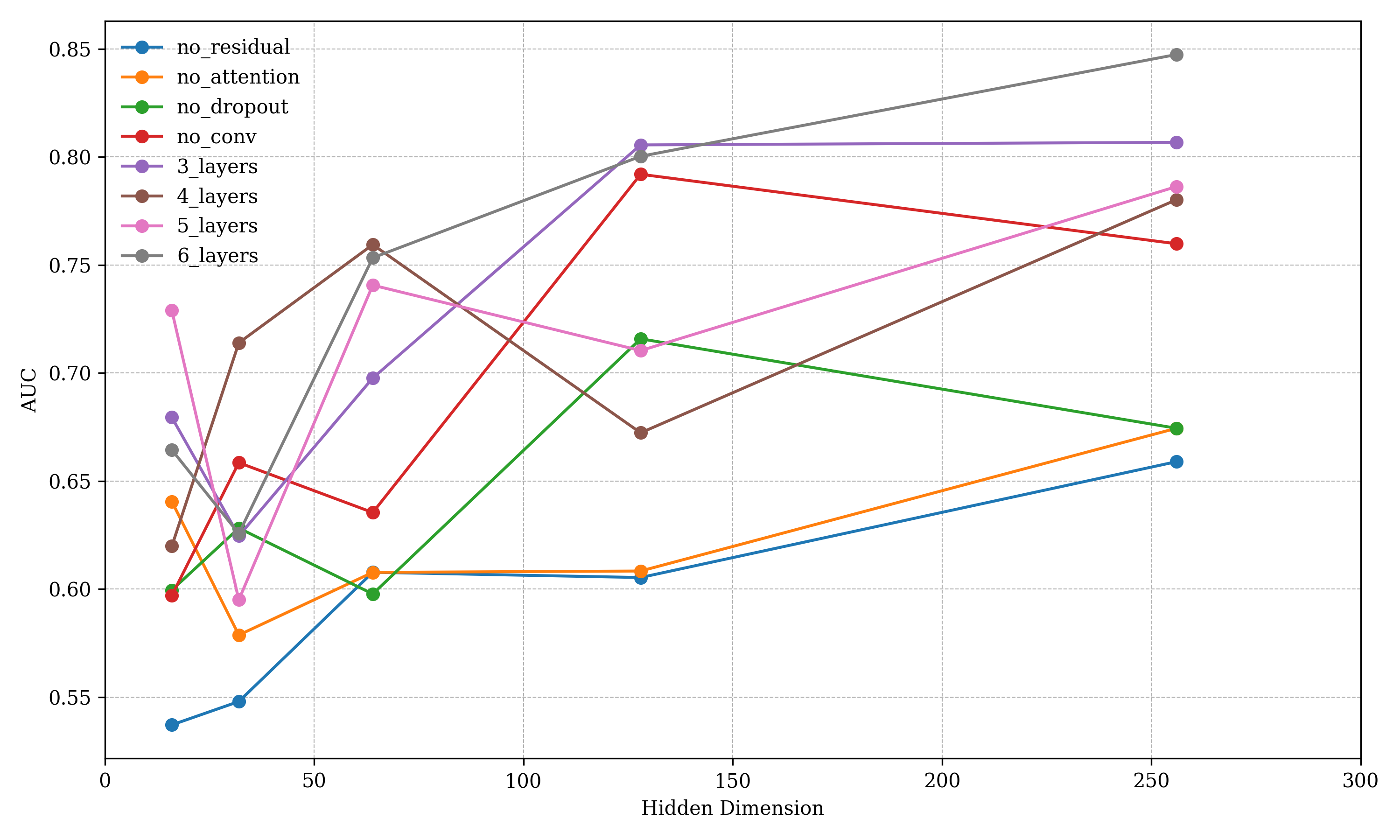}
  \caption{Comparison of seven experiments with different hidden dimensions in task 3.}
  \label{task3}
\end{figure}

\begin{figure}[htbp]
  \centering
  \includegraphics[width=0.5\textwidth]{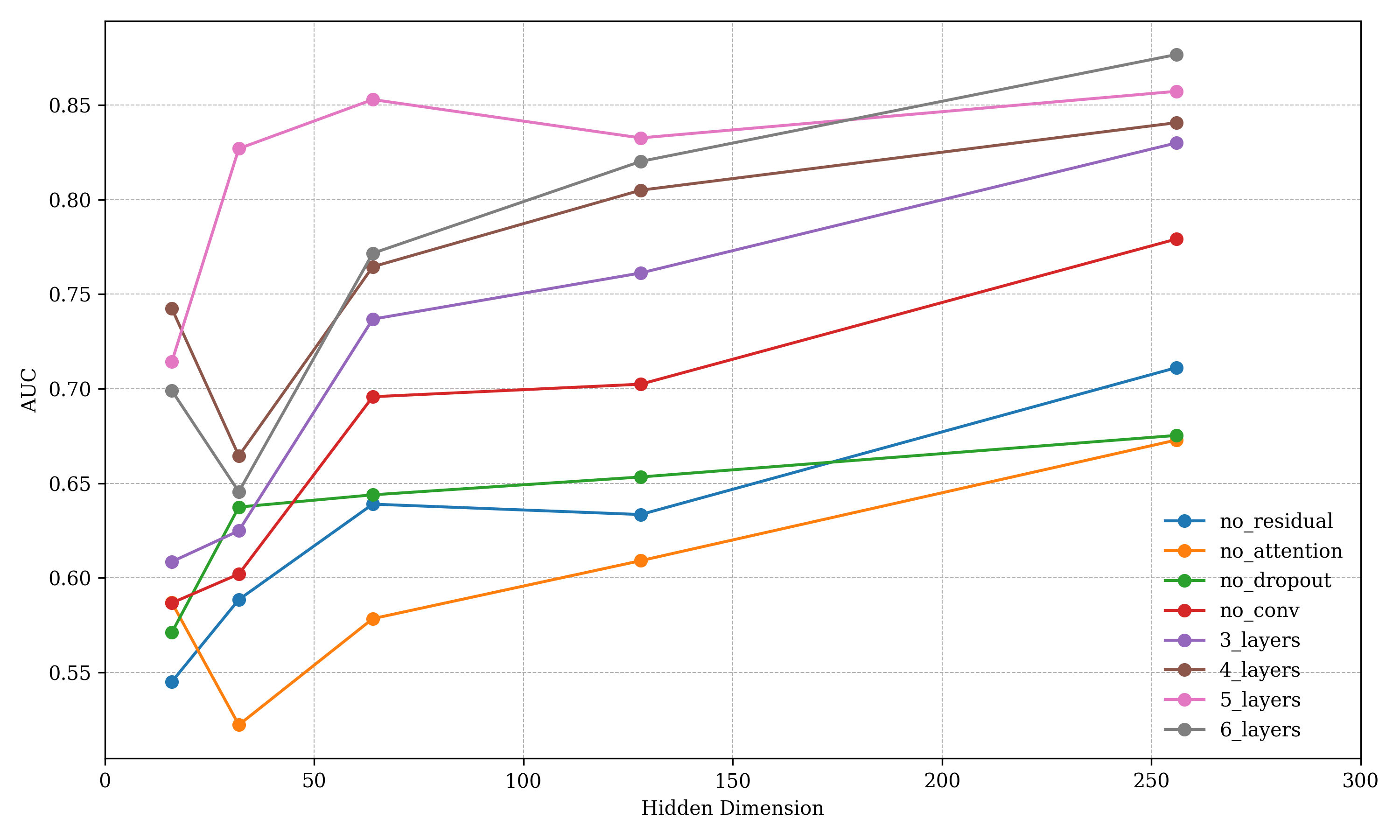}
  \caption{Comparison of seven experiments with different hidden dimensions in task 4.}
  \label{task4}
\end{figure}

\begin{figure}[htbp]
  \centering
  \includegraphics[width=0.5\textwidth]{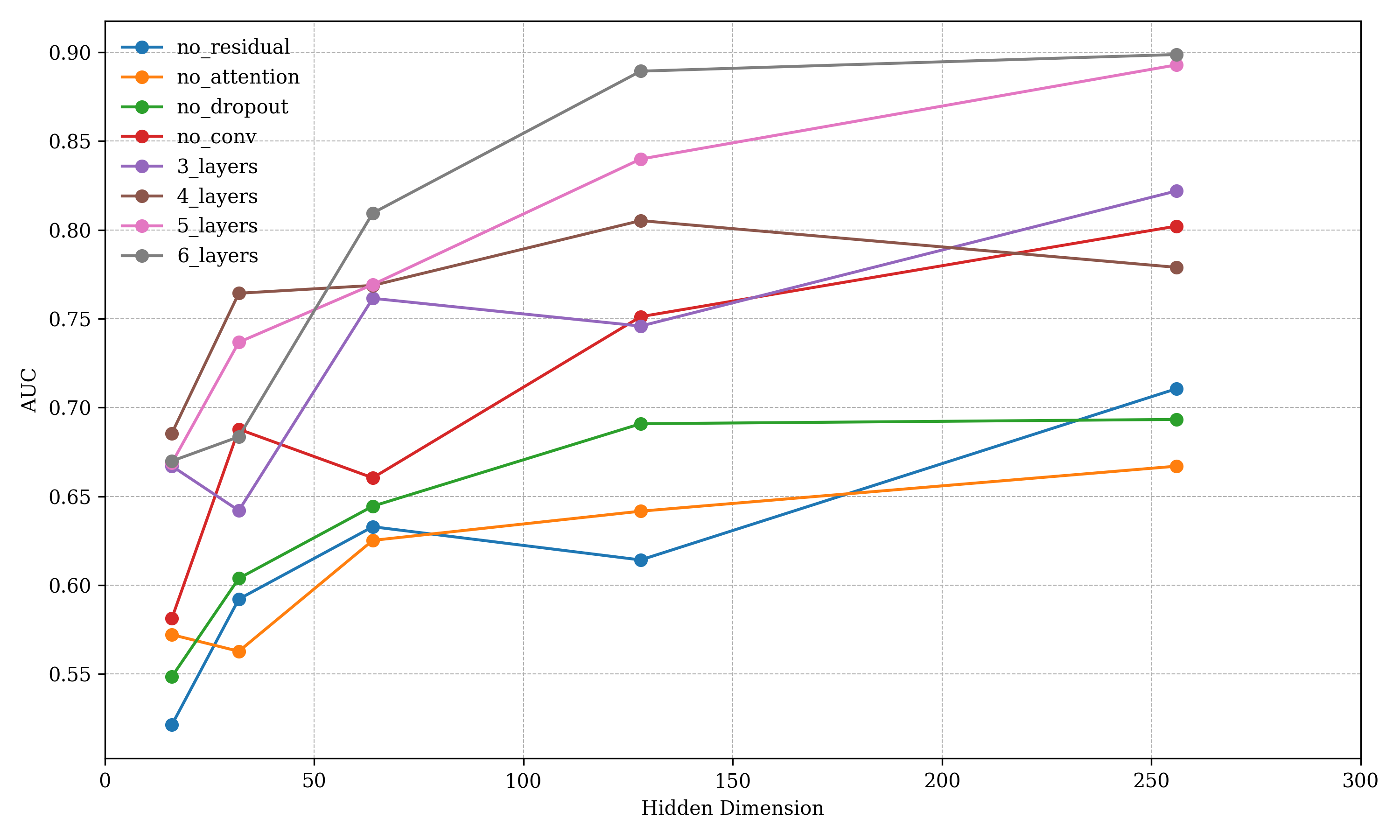}
  \caption{Comparison of seven experiments with different hidden dimensions in task 5.}
  \label{task5}
\end{figure}

\begin{figure}[htbp]
  \centering
  \includegraphics[width=0.5\textwidth]{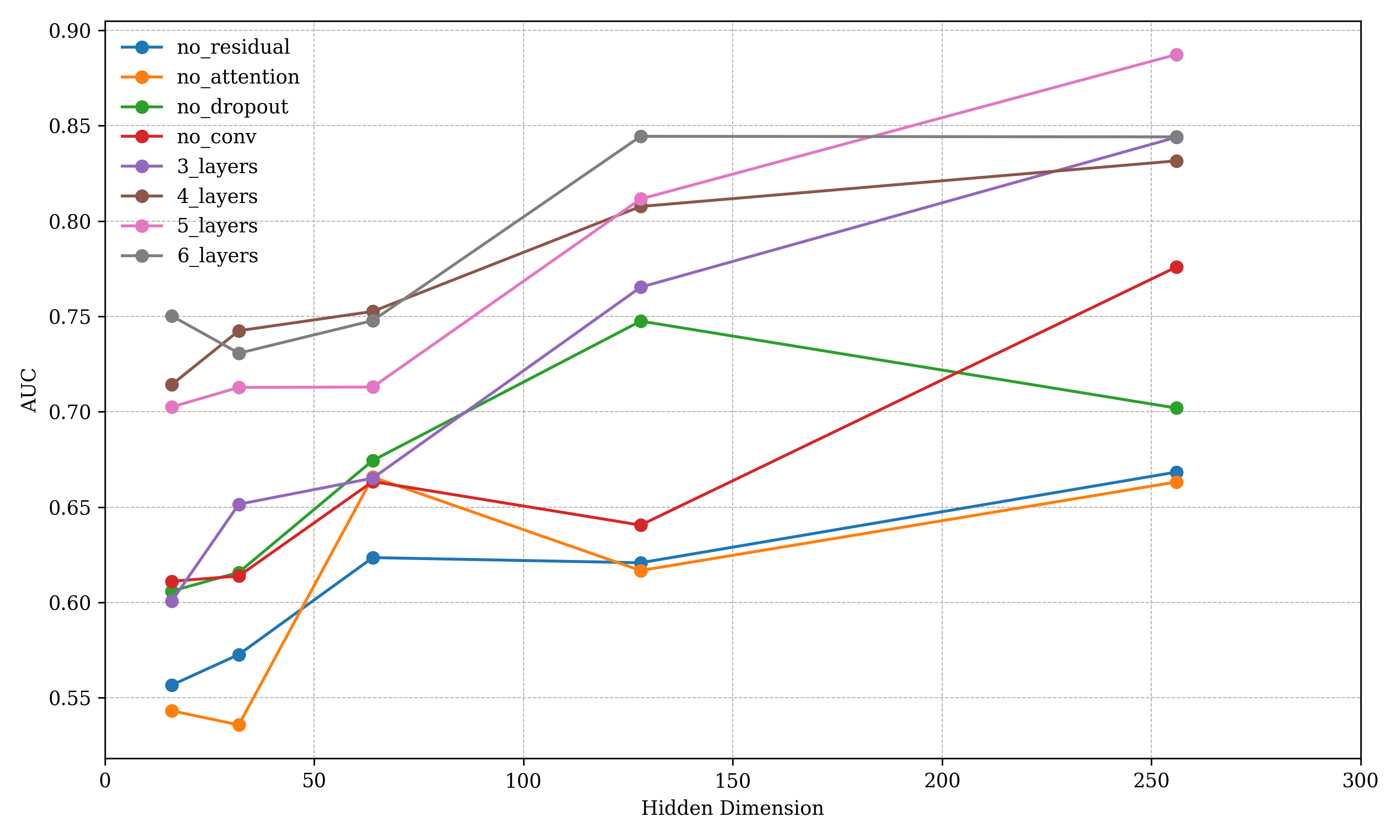}
  \caption{Comparison of seven experiments with different hidden dimensions in task 6.}
  \label{task6}
\end{figure}

\begin{figure}[htbp]
  \centering
  \includegraphics[width=0.5\textwidth]{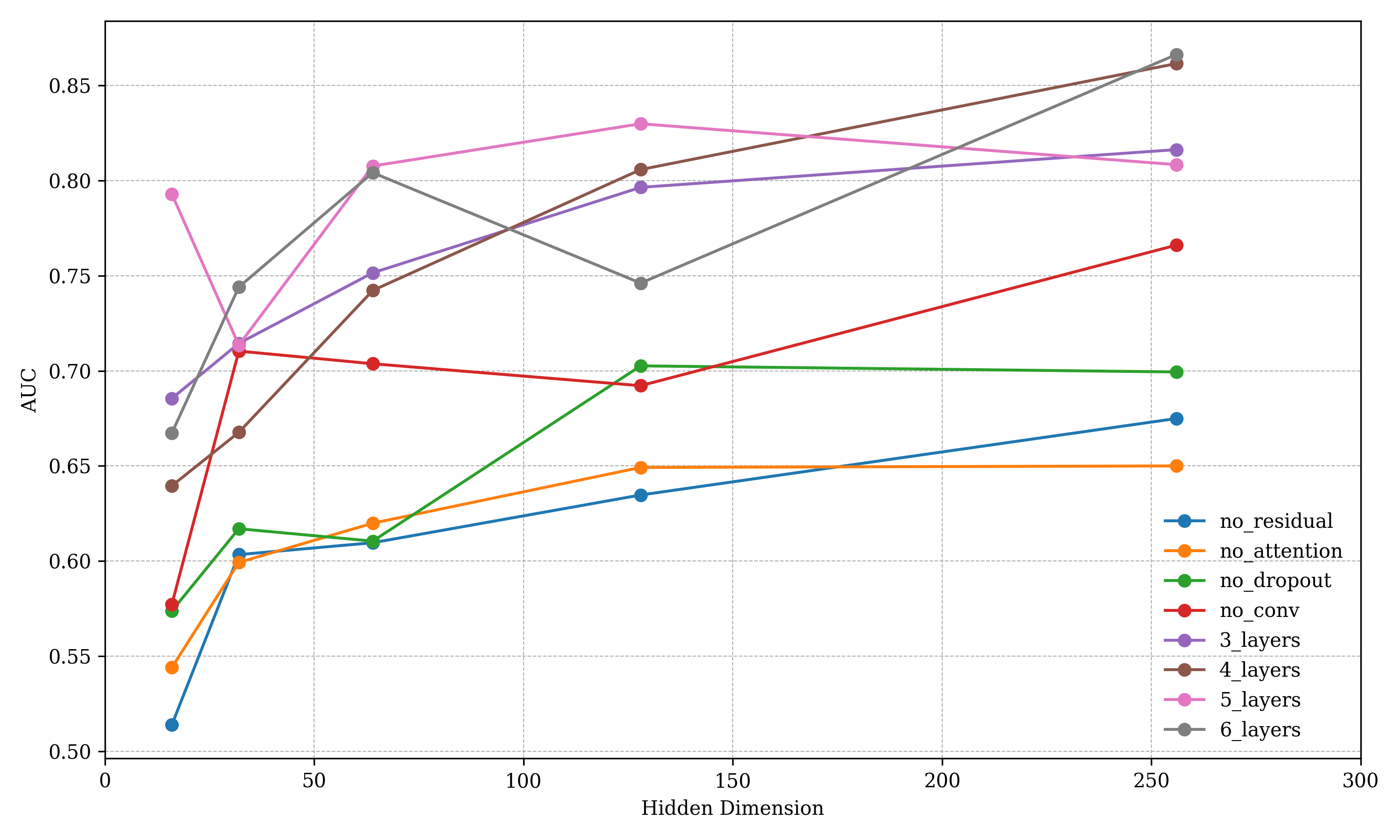}
  \caption{Comparison of seven experiments with different hidden dimensions in task 7.}
  \label{task7}
\end{figure}

\paragraph{Hidden Dimension}
The experimental results for the seven tasks are presented in Figures~\ref{task1} to \ref{task7}. In these figures, the horizontal axis represents the hidden layer dimensionality, systematically varied from 16 to 32 and then doubled up to 256 to examine its impact on diagnostic performance, while the vertical axis illustrates the corresponding performance metrics. Due to the relatively small size of our dataset, the experimental curves exhibit occasional fluctuations or "jumps." However, the overall trend aligns with our conclusions. These results provide a comprehensive analysis of how different architectural modifications affect the effectiveness of the proposed MFCN model. Notably, performance consistently improves as the hidden layer dimension increases across all datasets. For instance, in Task 1, across seven experiments, the model's performance improved from 0.4987 to 0.8892. Similar trends were observed in other tasks, confirming that increasing the hidden layer dimension enhances the model’s ability to capture complex features.

\paragraph{Components in the MFCN}
Furthermore, the results underscore the critical role of dropout, learnable attention layers, and residual connections in achieving high-quality outcomes. Dropout prevents overfitting by randomly deactivating neurons during training, introducing redundancy, and enhancing generalization. Learnable attention layers assign adaptive weights to information from different modalities, improving the model’s capacity to capture long-range dependencies and salient features, while convolutional layers facilitate the learnability of attention. Residual connections preserve original input information through skip connections, mitigate vanishing gradient issues in small-sample datasets, and enhance the stability and efficiency of deep network training. Additionally, increasing the number of ResNet layers further improves performance by enabling the model to learn more complex feature hierarchies. A deeper architecture facilitates the extraction of richer representations, leading to superior performance on challenging tasks and greater robustness in generalizing to new data.
\section{CONCLUSION}\label{VI}

In this paper, we propose a novel multimodal diagnostic framework, Medical Mimicry (MedMimic). This approach extends traditional dimensionality reduction by leveraging pre-trained large models to emulate experienced nuclear medicine physicians, transforming high-dimensional \(^{18}\)F-FDG PET/CT imaging data into semantically meaningful, low-dimensional feature tensors, thereby adapting to the complexity of the data. Furthermore, we introduce the Learnable Self-Attention Layer, designed to simulate the diagnostic reasoning of clinical physicians by assigning adaptive weights to information from different modalities, thereby enhancing the model’s diagnostic capability. Comprehensive experiments, including comparisons with single-modality models (both ML and DL) and ablation studies, validate the effectiveness of the proposed MedMimic framework across diverse tasks. By integrating the strengths of pre-trained large models and deep learning, MedMimic offers a novel perspective and an effective solution for disease classification tasks.

\section*{References}
\bibliographystyle{IEEEtran}
\bibliography{main}


\end{document}